\documentclass[sigconf]{acmart}

\setcopyright{acmlicensed}
\copyrightyear{2018}
\acmYear{2018}
\acmDOI{XXXXXXX.XXXXXXX}
\acmConference[Conference acronym 'XX]{Make sure to enter the correct
  conference title from your rights confirmation email}{June 03--05,
  2018}{Woodstock, NY}
\acmISBN{978-1-4503-XXXX-X/2018/06}
\usepackage{hyperref}
\usepackage{rotating}
\usepackage{multirow}
\usepackage{makecell}
\usepackage{array} 
\usepackage{subcaption}
\usepackage{framed}
\usepackage{enumitem}

\definecolor{deepred}{RGB}{139,0,0}

\newcommand\metric[1]{\textbf{#1}}
\newcommand\customquote[1]{``\textit{#1}''}
\newcommand\stat[1]{\textit{#1}}

\copyrightyear{2026}
\acmYear{2026}
\setcopyright{cc}
\setcctype{by-nc-nd}
\acmConference[CHI '26]{Proceedings of the 2026 CHI Conference on Human Factors in Computing Systems}{April 13--17, 2026}{Barcelona, Spain}
\acmBooktitle{Proceedings of the 2026 CHI Conference on Human Factors in Computing Systems (CHI '26), April 13--17, 2026, Barcelona, Spain}
\acmPrice{}
\acmDOI{10.1145/3772318.3790473}
\acmISBN{979-8-4007-2278-3/2026/04}
\begin{document}
\title{Playing the Imitation Game: How Perceived Generated Content Shapes Player Experience}

\author{Mahsa Bazzaz}
\affiliation{%
  \institution{Northeastern University}
  \city{Boston, Massachusetts}
  \country{USA}}
\email{bazzaz.ma@northeastern.edu}
\orcid{0009-0004-0022-9611}
\author{Seth Cooper}
\affiliation{%
  \institution{Northeastern University}
  \city{Boston, Massachusetts}
  \country{USA}}
\email{se.cooper@northeastern.edu}
\orcid{0000-0003-4504-0877}

\begin{abstract}
%

With the fast progress of generative AI in recent years, more games are integrating generated content, raising questions regarding how players perceive and respond to this content. To investigate, we ran a mixed-method survey on the games Super Mario Bros. and Sokoban, comparing procedurally generated levels and levels designed by humans to explore how perceptions of the creator relate to players' overall experience of gameplay. Players could not reliably identify the level's creator, yet their experiences were strongly linked to their beliefs about that creator rather than the actual truth. Levels believed to be human-made were rated as more fun and aesthetically pleasing. In contrast, those believed to be AI-generated were rated as more frustrating and challenging. This negative bias appeared spontaneously without knowing the levels' creator and often was based on unreliable cues of ``human-likeness.'' Our results underscore the importance of understanding perception biases when integrating generative systems into games.
\end{abstract}
\maketitle

\section{Introduction}


Procedural content generation (PCG) involves the automatic generation of any type of game content ranging from simple elements like textures and sprites to more complex components such as map layouts, game mechanics, game levels, and even full games~\cite{shaker_procedural_2016}. Procedurally generated content can empower novice designers and help solo developers with a limited budget to be able to finish their projects faster and more efficiently~\citep{boucher_resistance_2024}.

While PCG at its core often relies on fixed rules and algorithms to generate content, newer generative AI approaches create content from prompts using large models. Today, nearly 10,000 games\footnote{games with tag ``AI content'' in Steam DB}~\citep{steamdb_ai_content_disclose} on Steam use generative AI, and about 10,200 include procedural content\footnote{games with tag ``procedural content'' in Steam Store}~\citep{steam_store_procedural_generation}. This rapid adoption left many professional creators in games feeling simultaneously excited, overwhelmed, and concerned~\citep{vimpari_adapt_2023}. In January 2024, Steam introduced a new policy~\citep{steam_news_ai_policy} requiring developers to disclose how AI is used in their games. This raised the question of if transparency through labeling actually has the intended effect of helping players make more informed judgments.
\begin{figure*}[!h]
\centering
\includegraphics[width=0.9\textwidth]{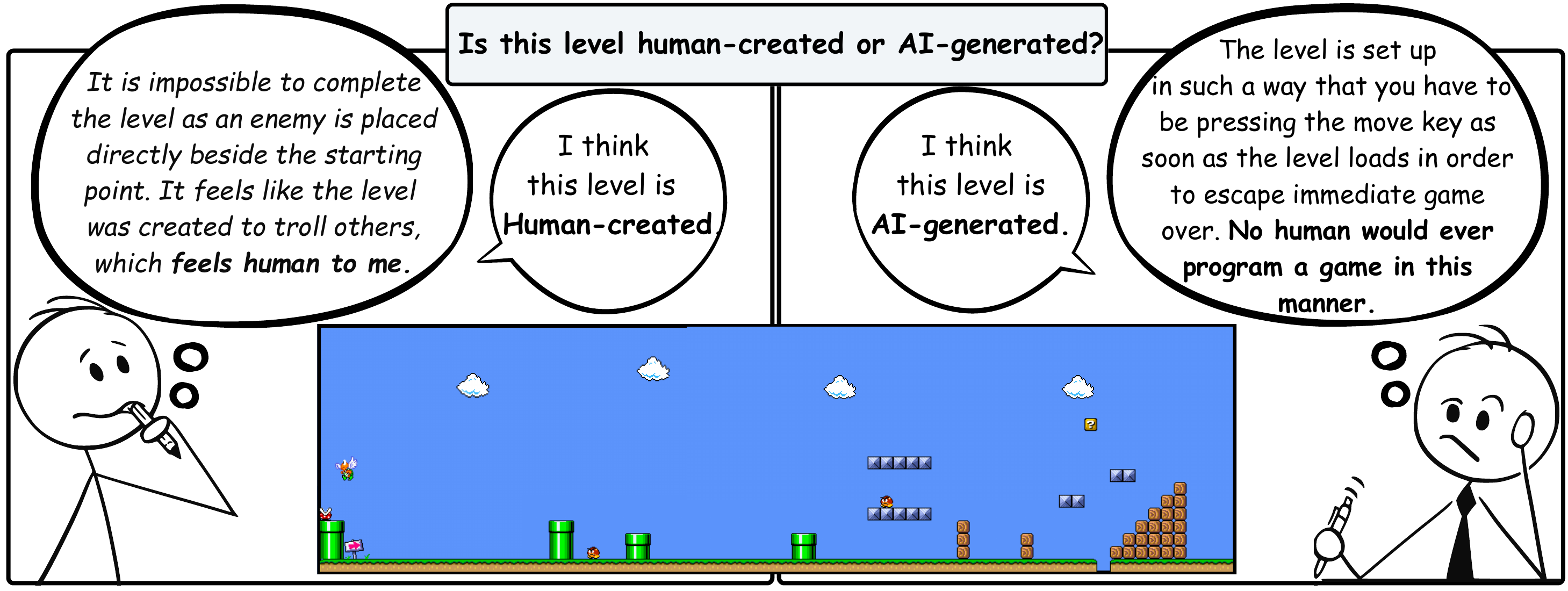}
\caption[Players have different perceptions about human-likeness.]{Players have different perceptions about human-likeness. The same game feature~(an enemy placed near the player's spawning position) was interpreted as evidence of AI design by one participant and as evidence of human design by another, illustrating how identical cues can support opposite judgments.\footnotemark}
\Description{A side-by-side comparison of two participants’ responses to the same Mario level. The first participant believes that the level is human-created, but the second participant thinks the opposite. As for the reason for their judgment, the same game feature~(an enemy placed near the player's spawning position) was interpreted as evidence of AI design by one participant and as evidence of human design by another, illustrating how identical cues can support opposite judgments.}
\label{fig:first_image}
\end{figure*}


The disclosure debate ties into a broader question: how do people perceive AI-generated content across different forms of media? Recent studies have approached the problem in two types of studies: First, studies in which they examine whether people can distinguish AI-generated content from human-created content, and how they try to do so. And second, priming studies in which they explore how labels and disclosure affect perceptions, including potential biases and placebo effects.

Building on the first type of studies, we would like to go beyond the ``Turing-like tests''~\citep{turing_computing_1950}, as it is one thing to recognize AI-generated content, and it is another if the player feels that this recognition impacts their experience. We shift the focus from the technical standpoint to a human perspective. We don't ask which procedurally generated content can blend in with human-created content and how. Instead, we ask how players' experience might change based on their perception of the content's creator. This question is specifically interesting when asked from different types of players with different stances on the acceptability of AI-generated content in games. In doing so, we change the question from ``\textit{Can you tell if it was made by AI?}'' to ``\textit{Does your perception of the creator---AI or human---change your experience of the level?}''

Unlike priming studies, which intentionally shape participants' expectations to study bias, our method instead simply observes their ``spontaneous judgments'' about the creator of the content and the association these judgments have with their experience. This observational design means we identify associations, not causal effects, but it provides insight into what occurs when platforms do not fully control disclosure, or when users draw incorrect conclusions based on their weak cues.

In this study, we try to answer five main research questions:
\vspace{2em}

\begin{framed}
\begin{itemize}[leftmargin=*, labelsep=0.5em]
\item \textbf{RQ1}: Can participants distinguish between levels created by humans and those created by AI?
\item \textbf{RQ2}: Does players' game playing background relate to their accuracy in distinguishing between AI- and human-created levels?
\item \textbf{RQ3}: What strategies do people use to differentiate between human-created and AI-generated levels?
\item \textbf{RQ4}: Is there a link between players' beliefs about the creator of the level and how they rate their experience with the game levels?
\item \textbf{RQ5}: Do players have different attitudes toward PCG and generative AI? How does this prior attitude relate to their reported experiences with the game levels?
\end{itemize}
\end{framed}

\vspace{0.5em}
Our study shows that players did not reliably distinguish between AI- and human-designed levels, yet their beliefs about creator were strongly linked with their reported experiences. The levels players believed to be human-created were rated as more fun and designed better aesthetically, while levels they believed to be AI-generated were more often described as frustrating and overly challenging. Players also expressed more negative attitude toward generative AI compared to PCG, and questioned its reliability, originality, and ethics. As AI-generated content becomes more and more common in games on platforms like Steam, understanding and addressing these perception-based biases is essential for both researchers and developers who aim to integrate AI in games responsibly and maintain trust in game communities.

The main contributions of this work are:
\begin{itemize}
    \item We show that players often misclassify AI- and human-created levels, and that their experiences align more with perceived creator than actual truth.  
    \item We show that players use fragile strategies to judge human-likeness, which can both guide and mislead their judgments.  
    \item We demonstrate that players hold different attitudes toward PCG and generative AI, shaped not only by gameplay and quality but also by broader ethical and value concerns, and that these prior attitudes are associated with their gameplay experiences.  
    \item We argue that perception bias poses challenges to trust, and we suggest that more nuanced disclosure practices are necessary as AI adoption increases in games.  
\end{itemize}
\footnotetext{Stick figures \copyright Zdenek Sasek via Canva.com.}

\section{Related Work}
\subsection{Games And Content Generation}\label{sec:pcg}
Game content generation is very different from other types of content, like images, because games are interactive experiences. This means the content has to follow strict rules to make sure players can actually complete the game~\citep{bazzaz_analysis_2025, yannakakis_procedural_2025}. For example, a platformer level might look nice, but if the level is blocked in some way and the player can’t reach the end, it’s not usable. Because of the playability concerns, game content generation faces different and often more complex challenges than other areas of generative systems~\citep{yannakakis_procedural_2025, summerville_procedural_2018}. 

Another important difference lies in how ``averageness'' is perceived in games. \citet{miller_ai_2023} found that AI-generated faces are often more average and less distinctive. This higher familiarity and lower memorability can sometimes make them seem ``more human'' than real human faces, creating a hyperrealism effect for images. In games, however, this same averageness can lead to what is called ``the 10,000 bowls of oatmeal problem''~\citep{compton_getting_2019}. Generated levels that are ``technically'' different, but they lack perceptual uniqueness, feel repetitive and cheaply made, and leave players with a less enjoyable experience. That's why literature on game content generation has created evaluation methods that measure expressiveness of the generators~\citep{smith_analyzing_2010, summerville_expanding_2018, withington_right_2023}, and distance metrics that quantify perceptual differences rather than technical differences~\citep{berns_not_2024}. 

These unique attributes make games a particularly interesting domain of study, which not only involves technical challenges but also includes strong emotional engagement from players~\citep{karpouzis_emotion_2016}, which places greater emphasis on how the experience is perceived.

\textbf{Procedural content generation~(PCG)} in games is the algorithmic creation of game content with limited or indirect user input \citep{shaker_procedural_2016}. Historically, PCG concepts originated in analog dice tables and dungeon generators and were designed to enhance replayability and support creative assistance in tabletop and early video games \citep{smith_analog_2015}. At its core, PCG aims to automate or assist content creation through formalized procedures (in contrast to hand-authored designs)~\citep{togelius_what_2011}. 

Classical procedural content generation methods include methods such as rule-based~\citep{shaker_constructive_2016} and search-based approaches~\citep{togelius_search_2010}. More recent PCG methods can be grouped into broad categories like constraint-based, reinforcement learning, and machine learning approaches~\citep{summerville_procedural_2018}. Constraint-based approaches generate levels by expressing constraints that describe what a valid solution looks like~\citep{font_constrained_2016, cooper_sturgeon_2022}. \textbf{PCG via Reinforcement learning (PCGRL)} includes approaches that use reinforcement learning approaches for content generation~\citep{yannakakis_procedural_2025}. While RL initially got popular for training playing agents~\citep{mnih_playing_2013}, it is now also a powerful PCG approach where level generation is framed as a sequential decision problem, with agents learning to place game elements through trial and error to maximize rewards~\citep{kaelbling_reinforcement_1996, khalifa_pcgrl_2020}. For the past decade, machine learning has been increasingly used for game content generation, introducing a new branch known as \textbf{PCG via machine learning (PCGML)}~\citep{summerville_procedural_2018}. These methods typically use neural networks trained on existing game content to generate new content that matches the style of the training data~\citep{summerville_procedural_2018,awiszus_toad_2020, volz_evolving_2018}. More recently, large language models~(LLMs) have emerged as a state-of-the-art approach in this area. LLMs' extensive pre-training on large amounts of data can be leveraged to generate game content like game levels~\citep{todd_level_2023,sudhakaran_mariogpt_2023}, game playing agents~\citep{joshi_towards_2025}, or narrative content~\citep{rist_sing_2024u} through natural language prompts. Unlike traditional PCGML approaches that require domain-specific training data, LLMs can leverage their general knowledge to generate content across multiple game domains~\citep{todd_level_2023, todd_benchmarking_2025}. However, this flexibility also comes with caveats of challenges around control, consistency, and ethics. Despite this rapid evolution of PCG methods from rule-based algorithms to modern generative AI there has been limited empirical investigation into how players perceive and experience content from these generation approaches, and whether their beliefs about a level's generation process would affect their gameplay experience.

\subsection{Turing Test}
Nearly a century ago, Alan Turing introduced the Imitation Game~\citep{turing_computing_1950} (now known as the Turing Test) as a way to ask whether a computer could exhibit behavior that is indistinguishable from human behavior. By having an interrogator exchanging written messages with both a person and a machine, Turing showed that if the machine's replies were convincingly mimicking human messages, we would have to admit its ``intelligence''. Over the years, different studies have debated whether this test captures intelligence or not~\citep{levesque_common_2017, hoffmann_ai_2022, schoenick_moving_2017, zador_toward_2022}. But ultimately, the Turing Test is a useful way to evaluate a machine's behavior from a black-box perspective.

With recent breakthroughs in AI, especially with LLMs, the Turing Test is more relevant than ever. As these generative models become more and more capable of producing high-quality content, recent work has suggested Turing-like tests for evaluating the model's ability to imitate human generative capabilities in different domains like texts (including poems \citep{kobis_artificial_2020, porter_ai-generated_2024}, stories \citep{clark_all_2021}, online reviews \citep{kovacs_turing_2024}, scientific abstracts \citep{nabata_evaluating_2025, else_abstracts_2023, hakam_human-written_2024}, and news articles \citep{bashardoust_comparing_2024, muda_people_2023, spitale_AI_2023, clark_all_2021}), images~\citep{meyer_find_2022, partadiredja_ai_2020, nightingale_ai-synthesized_2022, shen_study_2021, bray_testing_2023, lu_seeing_2023, ha_organic_2024, wang_human_2024}, videos~\citep{lewis_content_2022, angelides_breaking_2024, goh_understanding_2022}, audio~\citep{barrington_people_2025, mai_warning_2023, warren_better_2024, prudky_assessing_2023, muller_human_2022}, and visual arts~\citep{daniele_what_2021, vukojicic_imitation_2023, park_human_2024, samo_artificial_2023, wang_human_2024, gangadharbatla_role_2022, sun_pigments_2022, ha_organic_2024}. Typically, in these studies the Turing-like test includes showing the participants a set of content and asking them to decide whether each piece of content is AI or human generated, and to rate their confidence. 

Research on game-playing agents and believable AI frameworks~\citep{livingstone_turing_2006, hinkkanen_framework_2008} has also asked a similar question in this domain: \textit{can an agent play through a level in a way that mimics a human player?} Or similarly, \textit{can an algorithm create levels that resemble those designed by humans?} Therefore, Turing-like tests have been widely explored in games to evaluate how believable AI agents appear to players. In platformer contexts such as Super Mario Bros., this line of work was first formalized through the Mario AI competitions, which from 2009 to 2012 included a dedicated ``Turing Test Track'' where participants built character controllers intended to be judged on their human-likeness rather than pure performance~\citep{lee_learning_2014}. Similar studies on gameplay videos of 2D platformer agents \citep{shaker_turing_2013, camilleri_platformer_2016} and navigation tasks in 3D video game environments~\citep{devlin_navigation_2021, zuniga_how_2022, milani_navigates_2023} followed this line of work.

By 2015–2016, with growing attention on PCG following high-profile games like No Man's Sky~\citep{GAME_nomanssky} and its procedurally generated universes, an early attempt at a head-to-head comparison of procedurally generated game levels and hand-created levels was made. \citet{taylor_comparing_2015} studied player engagement in hand-crafted or procedurally generated Sokoban levels and found no significant difference in players' attention while playing hand-crafted and procedurally generated levels, showing that generated puzzles were at least as engaging as designer ones. Later on, studies have more directly applied the Turing test to evaluate the quality of the generated content, framing the quality as how well it can pass as human-created content~\citep{liu_automatic_2019, mourning_turing_2024, reis_human_2015}.

Building on similar Turing-like tests in previous work, we go beyond just asking whether players can tell AI- from human-made content, to examine how this perception itself relates to their gameplay experience.
\subsection{Expectation and Perception}
The placebo and nocebo effects provide a robust psychological framework for understanding how human expectations shape their experiences~\citep{colloca_placebo_2020}. The user's positive expectations can improve their experience and on the contrary the negative biases can overshadow actual quality, leading users to have a worse experience even if a product or system objectively performs well. In other words, what users believe about a system (accurate or not) can totally change how they feel about using it. 

Interestingly, previous work observed both of these two complementary forces in human perception of AI. Labeling something as AI can overshadow the actual quality. When people react positively to a stimulus under blind conditions, they may rate it worse once they think it's machine-made.  People judge news articles, health information, jokes, and other texts as lower quality when told they were AI-generated (even if they are not)~\citep{rae_effects_2024, porter_ai-generated_2024}. Although AI-generated artwork is sometimes preferred under blind conditions, it is often rated worse once told it's AI-generated (again, even if they are not)~\citep{grassini_understanding_2024, ragot_ai_2020, zhu_human_2024, gangadharbatla_role_2022}. 

On the other hand, it's been observed that minimal, even sham, AI framings can raise the perceived performance and enjoyment without changing the system at all. \citet{denisova_placebo_2015} told players a game included an adaptive difficulty when in fact it did not. But the players nonetheless reported higher immersion and enjoyment, suggesting that gameplay experience was shaped by players’ expectations. Likewise, in a word puzzle study, participants who believed they were receiving adaptive AI support (but they were not) solved more puzzles~\citep{kosch_placebo_2022}.

We take inspiration from these previous works studying the effects of perception on experience, but instead of using priming, we capture participants' spontaneous perceptions and how they connect to their experience. This better reflects how players form impressions in the wild when platforms do not provide informative labels to players.

\section{Methodology}
\subsection{Domain}
\setlength{\fboxrule}{0.2pt}
\setlength{\fboxsep}{0pt} 
\begin{figure*}[h]
    \centering
    \begin{subfigure}{0.81\textwidth}
        \centering
        \noindent\fbox{\includegraphics[width=\linewidth]{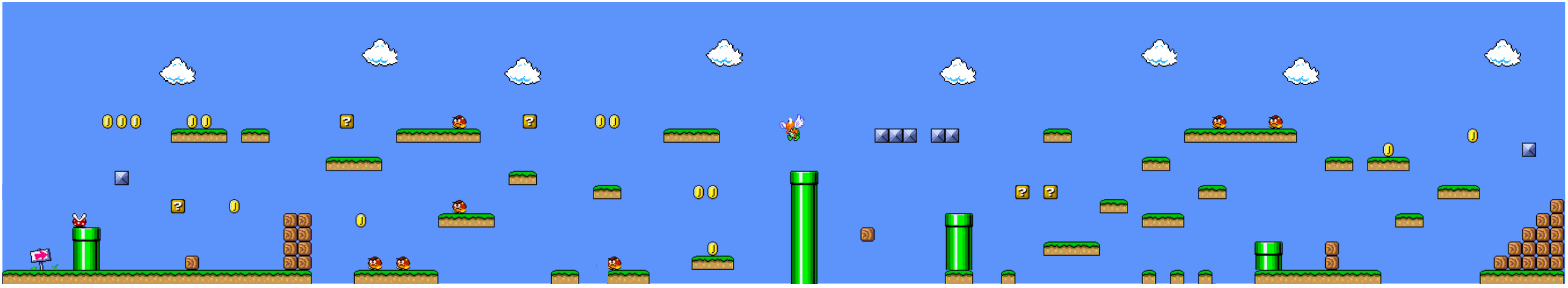}}
        \caption{A section of Super Mario Bros. level.}
        \Description{Screen shot of an example Super Mario Bros. level.}
        \label{fig:smb}
    \end{subfigure}
    \begin{subfigure}{0.149\textwidth}
        \centering
        \noindent\fbox{\includegraphics[width=\linewidth]{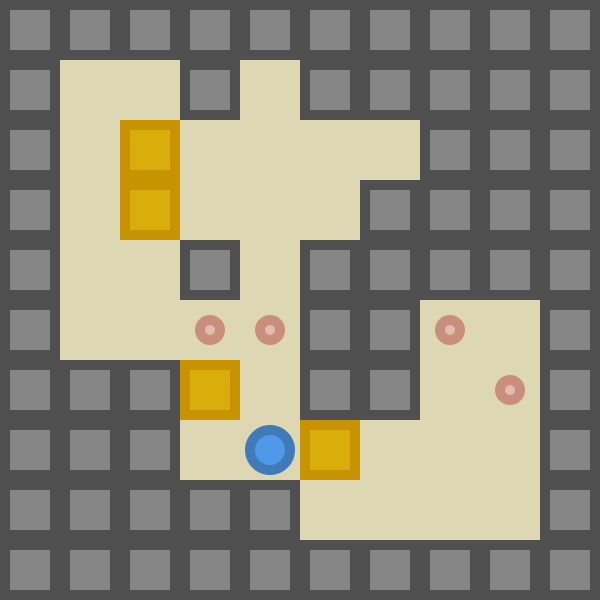}}
        \caption{A Sokoban level.}
        \Description{Screen shot of an example Sokoban level.}
        \label{fig:soko}
    \end{subfigure}
    \caption{Example screenshot of games.}
    \label{fig:game_examples}
\end{figure*}

2D tile-based game levels are the most common type of content in PCGML today~\citep{summerville_procedural_2018} with Mario-style platformer levels being used so often that they have become a standard dataset that functions as a benchmark for PCGML~\citep{summerville_procedural_2018}. 

Within 2D tile-based games, we chose Super Mario Bros. \citep{GAME_mariobros} and Sokoban \citep{GAME_sokoban} games as our domain of interest for two main reasons. First, they are established in the PCGML literature, with some of the earliest Turing-like tests conducted on them more than a decade ago~\citep{camilleri_platformer_2016, taylor_comparing_2015, reis_human_2015}. This provides us with a well-established basis for connecting our results to this body of work. Second, both games are simple and intuitive, requiring little onboarding for participants, which makes them especially suitable for an online crowdsourcing study where simpler tasks tend to produce higher accuracy~\citep{borromeo_influence_2016}. Finally, these two games come from two genres that are very different from each other along multiple dimensions like player goal, core mechanics, and cognitive load. Including these two games in our study gives us an opportunity to examine whether our results generalize across fundamentally different forms of play.

\textbf{Super Mario Bros.} is a platformer game created by Nintendo in which the player character jumps on platforms and enemies while avoiding their attacks and moving to the right of the scrolling screen to reach the exit~\citep{wiki_Super_Mario_2025}~(Figure \ref{fig:smb}).

\textbf{Sokoban}, created by Hiroyuki Imabayashi in 1981, is a puzzle game where the player is in a grid composed of floor squares and impassable wall squares and must push boxes onto marked storage locations~\citep{wiki_Sokoban_2025}~(Figure \ref{fig:soko}).

\subsection{Data Selection}
\subsubsection{\textbf{Human Authored Levels.}}
We collected all 15 original human-authored levels of Super Mario Bros. from the VGLC dataset \citep{summerville_vglc_2016}. The VGLC includes annotated level data for this platformer along with 11 other similar games. For Sokoban, we randomly (uniformly and without replacement) selected 15 levels from the 1,150 levels available in the dataset introduced by \citet{porsteinsson_sokoban_2009}, which contains levels created by different cited authors.
\subsubsection{\textbf{AI-generated Levels.}}
We chose our generation methods from a literature review on generation approaches in PCG to select a fair representation of the state-of-the-art generation methods. The complete detailed selection procedure is described in Appendix \ref{app:lit_rev}, and it resulted in the selection of 6 generation models (\citep{cooper_sturgeon_2022},~\citep{wang_fun_2022},~\citep{sudhakaran_mariogpt_2023}~\citep{cooper_sturgeon_2023}, ~\citep{racaniere_imagination_2017}, and ~\citep{todd_level_2023}) to be included in this study. The specific resources and the generation process for each approach are detailed in Appendix \ref{app:generation_resources}. This preserves the characteristics and limitations of each approach as originally published and avoids addition of confounding effects from our own implementation choices.

We detected two core challenges in selecting data for running the Turing test from previous work. The detected challenges are important because if not properly handled, they can undermine the reliability of Turing test results. 

\textbf{Core challenge 1}: \textit{The data selection procedure can be highly influential to the study results.} While running a Turing test on poems created with ChatGPT, \citet{kobis_artificial_2020} observed that when humans could handpick the best poems people were unable to reliably distinguish human from artificial poetry. However, when poems were randomly selected, people could detect the algorithm-generated poem with higher-than-chance levels.
    
\textbf{Core challenge 2}: \textit{AI-generated levels should not be readily distinguishable from human-created ones}~\citep{wang_human_2024}. When AI-generated content contains visible glitches or artifacts, it becomes trivially easy to identify without any close inspection. While one could argue that such flawed outputs represent a subset of generated content and should be included in evaluation, doing so may not yield any meaningful insights in the context of a Turing-like test.

To address the first core challenge, we avoided cherry-picking outputs from the generative models. Instead, we selected levels purely at random to eliminate any potential biases introduced by the manual selection. To tackle the second core challenge and make sure that the AI-generated content is not obviously distinguishable, we restricted the pool of randomly selected levels to those that are both ``playable'' and ``acceptable''. A level is considered ``playable'' if the player can reach the finish criterion of the level and is ``acceptable'' if there are no aesthetic flaws in the level \citep{bazzaz_analysis_2025}.  In other words, we selected levels that were free of obvious visual and functional flaws. For Mario, this meant verifying that all pipes were complete and for Sokoban, it meant ensuring that the number of goals matched the number of boxes, and there were no corner boxes (unmovable boxes). 

Based on these considerations, we gathered a pool of 100 levels generated with each generation system (or sampled from the original paper artifacts dataset) to choose five playable and acceptable random levels per method. This resulted in 15 AI-generated and 15 human-designed levels for each game.

This resulted in a pool of 60 unique game levels which varied based on \metric{Game} (Super Mario Bros. vs. Sokoban) and 
\metric{Creator} (AI-generated vs. Human-created). For each game, we 
included 15 human-created levels and 15 AI-generated levels. 
In total 30 levels per game and 30 levels per creator.

\subsection{Game Experience Metrics}
\label{sec:metric}
In order to study how players experience different game levels, we first need metrics that can capture those experiences in a meaningful way. Existing game experience questionnaires provide broad frameworks to assess player experience in full games, but many of their factors are not relevant when evaluating a single level's design. For example, the Game Experience Questionnaire (GEQ) core module measures seven dimensions including: Immersion, Tension, Competence, Flow, Positive/Negative Affect, and Challenge~\cite{poels_d3_2007}. These dimensions are built upon the assumption of a complete gameplay session. The Game User Experience Satisfaction Scale (GUESS) similarly covers nine subscales (Usability, Narrative, Visual/Audio Aesthetics, Creative Freedom, etc.) across $50+$ items~\citep{phan_development_2016}. The Player Experience Inventory (PXI) includes 10 constructs like Ease of Control, Audiovisual Appeal, Goals and Rules, and Progress Feedback (Functional Aspects), as well as Immersion, Autonomy, Mastery, Curiosity, and Meaning (Emotional Aspects)~\citep{abeele_development_2020}. These comprehensive scales are excellent for full games, but many dimensions (e.g. audio, controls, story) remain constant in the study of single game levels and won't be influenced by the level layout. In other words, asking about ``ease of control'' or ``audiovisual appeal'' makes little sense if only the level design (tile layout) changes. The same goes for deep immersion factors like ``narrative meaning'' or ``long-term progress feedback'', which don't apply to a single short level. The shortened versions of these scales, like the shortened version of the GUESS~\citep{keebler_validation_2020} and Mini Player Experience Inventory~\citep{haider_minipxi_2022} are also available, but they intend to fix the practical considerations where a longer measure is not feasible and still include the mentioned irrelevant dimensions to shorter game level experiences. 

Therefore, it's important to tailor the questionnaire to focus on experiential dimensions that a level's design can impact. Prior research on level evaluation and player experience in PCG has identified experiential measures that are suitable for judging procedurally generated game levels that include measures of players' short experience with the game levels rather than using a full game UX battery \citep{guzdial_game_2016, pedersen_modeling_2009, guzdial_conceptual_2021, guzdial_friend_2019}. 

Here we provide the list of the factors extracted from previous work for the evaluation of game levels based on their level design, and without the full gameplay factors:
\begin{itemize}
    \item \textbf{Enjoyment / Fun}: This is a fundamental measure of how much the participant enjoyed the level. Fun or enjoyment is frequently assessed in level evaluation studies~\citep{guzdial_game_2016, pedersen_modeling_2009, guzdial_deep_2021, guzdial_learning_2016, guzdial_friend_2019, summerville_understanding_2017}. This metric is well-supported in prior work (it aligns with the “Positive Affect” aspect of GEQ). Participants rated this dimension using the statement: ``It was fun to play.''
    
    \item \textbf{Difficulty / Challenge}: Perceived challenge is another key dimension, reflecting how difficult or easy the level felt. Level layout directly affects challenge (through enemy placement, puzzle complexity, etc.), so it's crucial to measure. Challenge appears in almost every game experience model (GEQ core scales and part of PXI's functional constructs) and many studies include this metric \citep{pedersen_modeling_2009, guzdial_conceptual_2021, biemer_solution_2024, guzdial_deep_2021, guzdial_learning_2016, guzdial_friend_2019, summerville_understanding_2017}. For this metric, participants responded to the item: ``It was very challenging.''

    \item \textbf{Frustration / Negative Affect}: Frustration captures the negative side of the experience because it typically arises if the level was unfair, too difficult or without reward. Prior work treats frustration as a distinct, important metric. In general game UX research, frustration maps to the Negative Affect/Tension dimension (in GEQ). This metric has been used in multiple player experience studies as a key outcome to minimize \citep{pedersen_modeling_2009, yu_personalized_2011, guzdial_conceptual_2021, guzdial_learning_2016, guzdial_friend_2019}. This item was worded as: ``I felt frustrated while playing it.''
    
    \item \textbf{Novelty / Surprise}: Novelty measures how original, surprising, or innovative the level felt. This metric is supported by PCG literature \citep{chakraborttii_towards_2024, pedersen_modeling_2009, guzdial_conceptual_2021, guzdial_friend_2019, anjum_ink_2024}.  We presented this metric using the statement: ``It had surprises or unexpected elements.''
    
    \item \textbf{Aesthetics / Design Quality}: This asks the player to evaluate how well-designed the visual presentation is. Prior work has used similar constructs \citep{guzdial_game_2016, guzdial_conceptual_2021, guzdial_deep_2021, guzdial_learning_2016, anjum_ink_2024, summerville_understanding_2017}. Participants rated this metric with the statement: ``The layout felt well-designed and high quality.''
\end{itemize}

Although these items do not come from an established framework, each directly targets a separate well-supported experiential construct. Altogether, they form a focused and intuitive set of measures that show face validity~\citep{holden_face_2010} as they directly map onto experiential dimensions that are relevant in short-term level play. Each metric was assessed using a single-item measure on a 5-point scale with anchors: 1 = Strongly disagree, 2 = Somewhat disagree, 3 = Neither agree nor disagree, 4 = Somewhat agree, 5 = Strongly agree.
\subsection{Participants Recruitment}
We recruited a sample of 154 participants via Prolific. This sample size was typical in similar mixed-effects analyses~\citep{meyer_find_2022, shen_study_2021, lu_seeing_2023, bray_testing_2023} and provided us adequate power to detect meaningful effects. Pre-screening criteria required that all participants be above 18, be in the United States and know English (inclusion criteria). 

Based on Prolific's demographic data, the participants included 75 males, 77 females, and 2 individuals who did not disclose their gender. The participants were between 21 and 72 years old, with average of 42.73 years old~(\stat{SD} = $10.81$). Inspection of Q–Q plot indicated no major deviations from normality. Each Prolific user who signed up for our survey was redirected by a link to our survey on Qualtrics. Participants were compensated at an average rate of \$8.87/hour, above Prolific’s \$8 minimum pay rate. The base payment for completing the survey was calibrated for \$8/hour, and participants received extra \$0.10 bonuses for each qualitative response. Such financial incentives are common in behavioral research to improve accuracy and reduce measurement error \citep{kobis_artificial_2020, ariely_psychology_2007, schlag_penny_2015}. This study protocol was reviewed and approved by the authors' Institutional Review Board (IRB).
\subsection{Procedure}\label{sec:procedure}
Each participant was presented with 6 levels that were sampled randomly without replacement from the pool of 60 unique levels. The sampling was balanced across the four categories of \metric{Game} and \metric{Creator} at the population level, meaning that while individual participants did not see all games and creators, the aggregate data provides balanced coverage across all 60 unique game levels.

With 154 responses to our survey, after excluding the unfinished surveys, there were 142 valid responses, each including 6 trials (852 in total). This means each of the 60 unique game levels was evaluated by approximately 14 participants ($\frac{852}{60} \approx 14.2$), providing multiple independent judgments per unique game level, while keeping individual participation time manageable (\textasciitilde 18 minutes).

Participants started by reading a description of the survey and a consent form, and were asked to continue if they understood and gave consent. Previous work conducting Turing-like tests that included targeted tutorials before the study and feedback on participants' performance during the study, reported a very small 10–15\% increase in accuracy \citep{bray_testing_2023, mai_warning_2023} or even no improvement \citep{muller_human_2022}. Therefore, our participants were not provided with education or training on how to identify AI-generated content before participating in the study, and the correct answers were not given upon completion to avoid any learning effect.  Instead, in order to make sure all participants had the minimum baseline knowledge about the games, at the beginning of the survey, we included a short description of each game explaining the setup and the goal in levels and asked if they had played that or a similar game.

After this introduction, participants were presented with the 6 trials. Each trial included a playable level that they could interact with and play and a set of follow-up questions regarding that level. Firstly, a two-alternative forced-choice (2AFC) question would ask participants, ``Is this level AI-generated or human-designed?'' and a 3-point Likert scale to report their confidence in their choice with~(``Not Confident'', ``Somewhat Confident'', ``Confident''). Then participants had to rate their experience with the level based on the five dimensions discussed in section \ref{sec:metric}

Lastly, for each level participants would answer an open-ended question on the reason they made their choice. This part of the survey directly addresses RQ3 and also supports the validation of RQ1. Prior research highlights the importance of understanding participants' reasoning, as their implicit evaluation criteria may not align with those intended by the researchers \citep{clark_all_2021, warren_better_2024}. A screenshot of the survey as it appeared to one participant is provided in Figure \ref{fig:survey}.

Following all the trials, we asked participants whether they felt differently about applications of PCG and generative AI in games. In order to make sure that participants had a baseline understanding of PCG and generative AI, we included very simple examples of these methods (Minecraft, No Man's Sky, and Spelunky for PCG and ChatGPT and Midjourney for generative AI). If they did, we asked them about their view on PCG and generative AI separately using two 5-point Likert scales (-2=Very negative, -1=Negative, 0=Neutral, 1=Positive, 2=Very positive). Then we asked them to elaborate on their views in an open-ended response to get a better understanding of the roots of the opinions. Those who did not indicate a difference were recorded as neutral view.
Lastly, we asked participants background questions about their general gaming habits (such as how often they play and whether they play action-adventure or puzzle games), and their experience with game design. This information may help us analyze whether prior familiarity and domain knowledge significantly influence participants' performance, as was observed in the visual art domains in which artists were able to be 87\% accurate in their areas of expertise, while general audiences remained close to chance~\citep{ha_organic_2024}.

As part of our quality assurance process, we checked if any individual's average response time fell more than three standard deviations below the mean (indicating inattentive or automated responses). 2 participants' responses fell into this criterion and were not included in the analysis. This resulted in eliminating 12 trials out of the 852 trials retrieved (1.4\% elimination). The timer was hidden from participants, and they were not instructed to pay attention to timing at all.
\subsection{User Study Evaluation}
\begin{figure*}[h]
    \centering
    \begin{minipage}{0.45\textwidth}
        \centering
        \includegraphics[width=\linewidth]
        {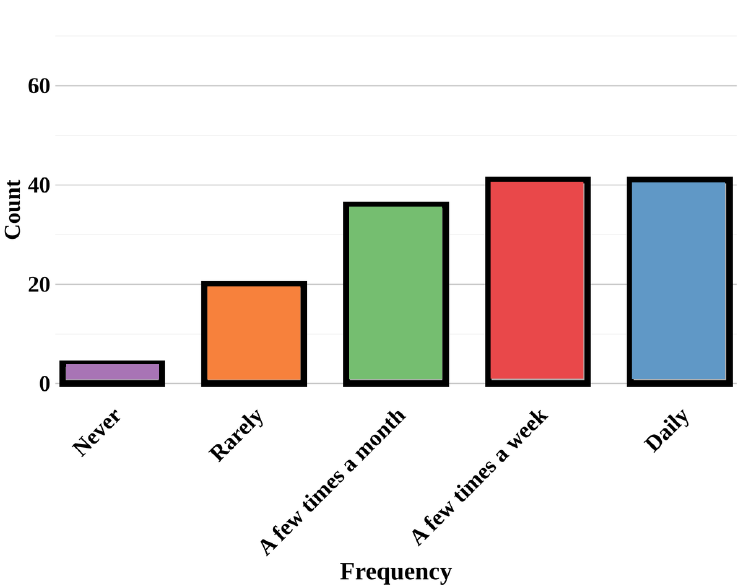}
        \caption{Participants' gaming frequency.}
        \Description{Bar chart showing participants' gaming frequency. Most play either daily or a few times a week, with fewer playing monthly, rarely, or never.}
        \label{fig:game_play}   
    \end{minipage}\hspace{0.01\textwidth}
    \begin{minipage}{0.45\textwidth}
        \centering
        \includegraphics[width=\linewidth]
        {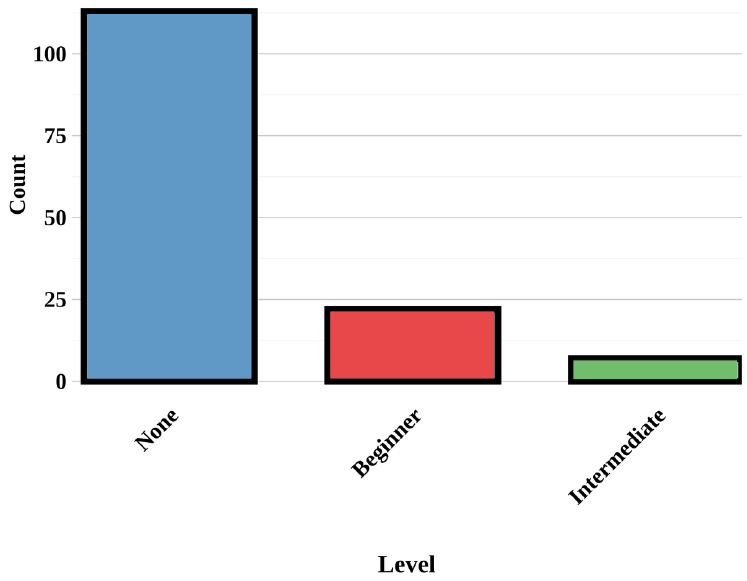}
        \caption{Participants' experience with game design.}
        \Description{Bar chart showing participants' game design experience levels. Most have no experience, fewer are beginners, and very few are at an intermediate level.}
        \label{fig:game_design}
    \end{minipage}

    \vspace{0.5cm} 

    \begin{minipage}{0.9\textwidth}
        \centering
        \includegraphics[width=\linewidth]{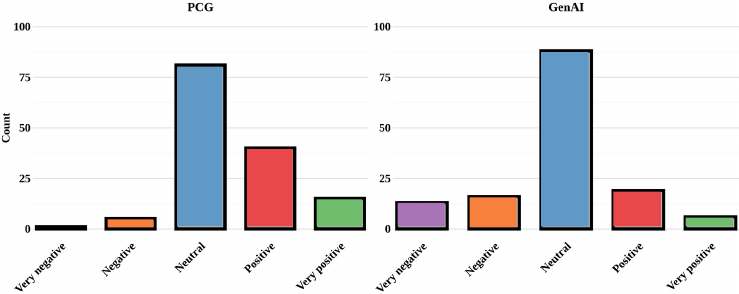}
        \caption{Participants' attitudes toward procedural content generation (PCG) and generative AI (GenAI) in games. Bar charts show the distribution of responses across five categories from ``Very negative'' to ``Very positive.''}
        \Description{Side-by-side bar charts showing participant attitudes toward PCG and GenAI in games across five sentiment categories. Both charts peak at ``Neutral,'' with PCG having more ``Positive'' and GenAI more ``Very negative'' responses.}
        \label{fig:bias}        
    \end{minipage}\hfill

\end{figure*}

We quantified participants' ability to distinguish AI- from human-created levels using different measures of accuracy borrowed from previous work~\citep{ha_organic_2024, warren_better_2024}:
\begin{itemize}
    \item Overall accuracy.
    \item False Positive Rate (FPR) indicating the player wrongly labeled a real human level as AI.
    \item False Negative Rate (FNR) indicating the player wrongly labeled an AI-generated level as human.
    \item AI Detection Success Rate (ADSR).
\end{itemize}
To test whether accuracy exceeded chance, we ran a two-sided binomial test on the proportion of correct responses (i.e., cases where the participant's belief about a level's creator matched the truth) against \stat{$p_0=0.5$}.

To examine whether participants' background in game playing frequency relates to correctness (belief = truth), we use fixed-effects logistic regression and the true source of creation and game as covariates in the model. We also examined the relation between the confidence in the decision and correctness with the same approach. 

Players' experience metrics were collected based on a $-2$ to $+2$ Likert scale. Because each player rated many items and each unique game level was seen by many participants, ratings are repeated and crossed. Therefore, we used ordinal logistic regression with random intercepts for participant and unique game level. Likelihood-ratio tests (CLMM vs. CLM without random effects) showed that including random effects significantly improved fit across outcomes.

For each player experience metric, we analyzed two sets of models. First we model with predictors belief, truth, and game to analyze the relationship between players' beliefs on the source of creation of the level and their experience. Second, we model with predictors attitudes toward PCG and toward generative AI, plus truth and game to analyze the relationship between previous biases toward AI and player experience with levels. 
\begin{figure*}[h]
    \centering
    \begin{minipage}{0.45\textwidth}
        \centering
        \includegraphics[width=\linewidth]{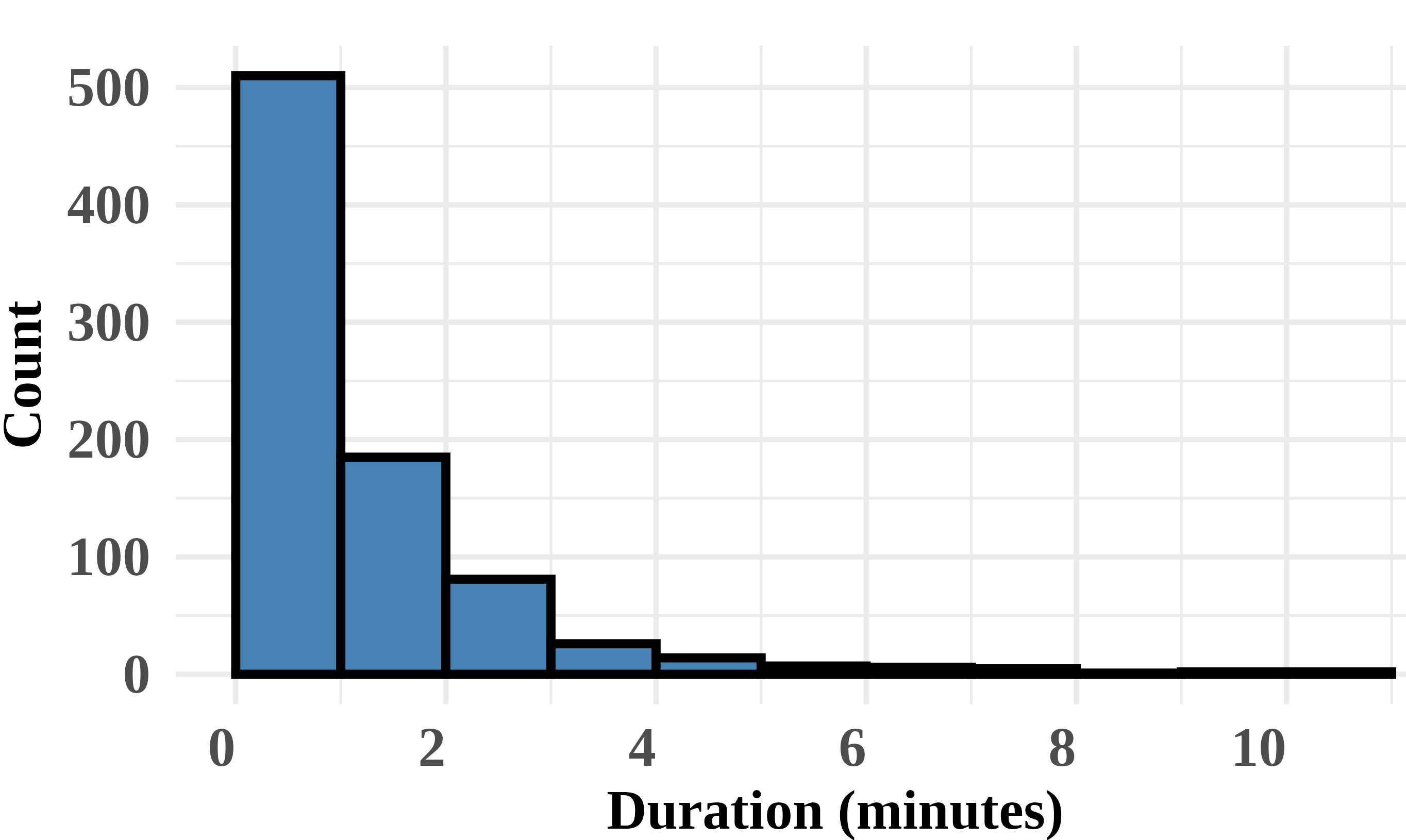}
        \caption{Amount of time participants spent playing the game levels. Participants spent an average of 1 minute 7 seconds per level, with a median of 42 seconds (min = 0 seconds; max = 11 minutes 24 seconds).}
        \Description{Amount of time participants spent playing the game levels. Participants spent an average of 1 minute 7 seconds per level, with a median of 42 seconds (min = 0 seconds; max = 11 minutes 24 seconds).}
        \label{fig:time}   
    \end{minipage}\hspace{0.01\textwidth}
    \begin{minipage}{0.45\textwidth}
        \centering
        \includegraphics[width=\linewidth]{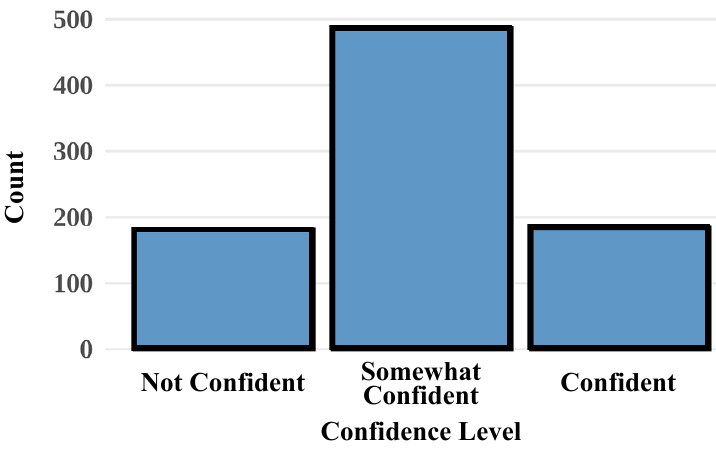}
        \caption{Distribution of participants’ reported confidence in their decisions. Most decisions were made with `somewhat confident' responses (n = 484), with substantially fewer rated as `not confident' (n = 181) or `confident' (n = 175).}
        \Description{Distribution of participants’ reported confidence in their decisions. Most decisions were made with `somewhat confident' responses (n = 484), with substantially fewer rated as `not confident' (n = 181) or `confident' (n = 175).}
        \label{fig:confidence}
    \end{minipage}
\end{figure*}
\subsection{Thematic Analysis}
We proceeded with a thematic analysis~\citep{braun_using_2006, terry_thematic_2017} to find patterns and meaning in our qualitative data. Our study included two types of open-ended responses: (1) players' reasoning for classifying levels as AI- or human-created and (2) background questions about their views on generative AI and PCG. For each section, we conducted a separate coding process and thematic analysis following the same process.

To build the coding scheme, the first author first reviewed all responses from both sections and assigned initial categories. These categories were then refined through multiple passes to ensure they were distinct and relevant to our research goals.

To check inter-rater reliability~\citep{hallgren_computing_2012}, a second coder was trained on the categories and independently coded the responses. We then calculated Cohen's kappa~\citep{cohen_statistical_2013} to assess agreement. We treated each full sentence as the unit of coding, and we assigned 1 to 3 codes to each unit. Each rater also had the option to mark a response as ``Bad Response'' if they believed the response did not appropriately answer the question. We removed all responses that were labeled as bad responses. 

For reasoning responses, the average Cohen's kappa across all codes was 0.76 (\stat{SD = 0.12, range = 0.54–0.89}) with an average agreement of 96.78\%, indicating a moderate level of agreement~\citep{mchugh_interrater_2012}. For background responses about generative AI and PCG, the average Cohen's kappa was 0.80 (\stat{SD = 0.12, range = 0.66–1.00}) with an average agreement of 97.92\%, indicating a strong level of agreement~\citep{mchugh_interrater_2012}.

Following coding, we grouped related codes into themes based on similarity and conceptual coherence. We used NVivo~\citep{NVivo14} for all these processes.

\section{Findings}
\subsection{RQ1: Can participants distinguish between levels created by humans and those created by AI?}\label{sec:RQ1}
Participants identified the true creator correctly on 53\% of trials (430 out of 812). To ensure that 50\% corresponded to exact chance performance, we balanced the number of trials by removing samples from the larger group, resulting in equal numbers of trials with human-curated and AI-generated levels (406 trials per group). Under the null hypothesis that accuracy would not differ from chance (50\%), a two-sided binomial test showed no significantly deviation (\stat{p = .099, 95\% CI [49.5\%, 56.4\%]}). The false positive rate 26.6\% and false negative rate 26.3\% suggests that participants were equally likely to misclassify human levels as AI as they were to misclassify AI levels as human. The AI detection success rate 47.2\% shows that participants were correct in identifying AI levels less than half the time, again consistent with chance performance.

The omnibus likelihood-ratio tests (type III ANOVA) showed that being confident in the decision was not a significant predictor of correctly identifying the source of creation (\stat{$\chi^2(2)$ = 1.35, p = .51}). Looking into confidence ratings themselves~(Figure \ref{fig:confidence}), ordinal logistic regression showed a significant effect of gaming frequency on confidence ratings (\stat{$\chi^2(4)$ = 12.58, p = .014}), with more frequent players reporting greater confidence in their decisions (\stat{$\beta$ = 0.72, SE = 0.28, p = .011}). There was weak evidence that participants felt less confident in deciding the true origin of levels when they were judging Sokoban levels compared to Mario levels (\stat{$\chi^2(1)$ = 3.35, p = .067}).
\subsection{RQ2: Does players' game playing background relate to their accuracy in distinguishing between AI- and human-created levels?}\label{sec:RQ2}
We tested whether participants' self-reported gaming frequency (Figure \ref{fig:game_play}) was related to their accuracy in detecting AI-generated game levels. A logistic regression with predictors gaming background, true creator, and game showed a main effect of gaming background, (\stat{$\chi^2(4)$ = 12.49, p = .014}). Neither the true creator (\stat{$\chi^2(1)$ = 0.33, p = .57}) nor the game, (\stat{$\chi^2(1)$ = 0.09, p = .77}) were significant predictors.

Accuracy did not increase monotonically with more frequent gaming though. Instead, participants who self-reported playing games rarely performed better in finding which levels are human-created and which are AI-generated than those who reported playing weekly or even daily. This result suggests that playing more games doesn't necessarily make players better at distinguishing between AI-generated and human-created content and it may even bias them toward heuristics that are not accurate. Following the same pattern, prior experience with level design~(Figure \ref{fig:game_design}) also did not help players be more accurate ($\chi^2$(2) = 0.99, $p = .61$). However, because of the uneven distribution of experience levels in our distribution, the statistical power to detect effects among participants with higher design experience in our analysis is limited.

We also checked whether knowing the games or liking the genres helped participants do better. For Mario, having played the game before (n=137) did not improve accuracy (\stat{$\beta$ = –0.10, SE = 0.59, p = .864}), and people who said they play action-adventure games (Super Mario Bros. genre, n=98) were also not more accurate (\stat{$\beta$ = –0.18, SE = 0.15, p = .21}). For Sokoban, having played the game (n=88) showed a small but non-significant positive effect (\stat{$\beta$ = 0.17, SE = 0.20, p = .39}), and puzzle players (Sokoban genre, n=120) were not more accurate either (\stat{$\beta$ = 0.23, SE = 0.19, p = .22}). 

This further suggests that frequent players may bring strong expectations about how a Mario or Sokoban level ``should'' look, which can lead them to misclassify AI-generated content. By contrast, participants who rarely play games approach the task with fewer preconceptions. This interpretation also matches our confidence analyses in Section \ref{sec:RQ1} in which we observed that frequent players felt more confident in their answers, but they were not actually more accurate.
\subsection{RQ3: What strategies do people use to differentiate between human-created and AI-generated images?}
\begin{figure*}[!h]
\centering
\includegraphics[width=0.95\textwidth]{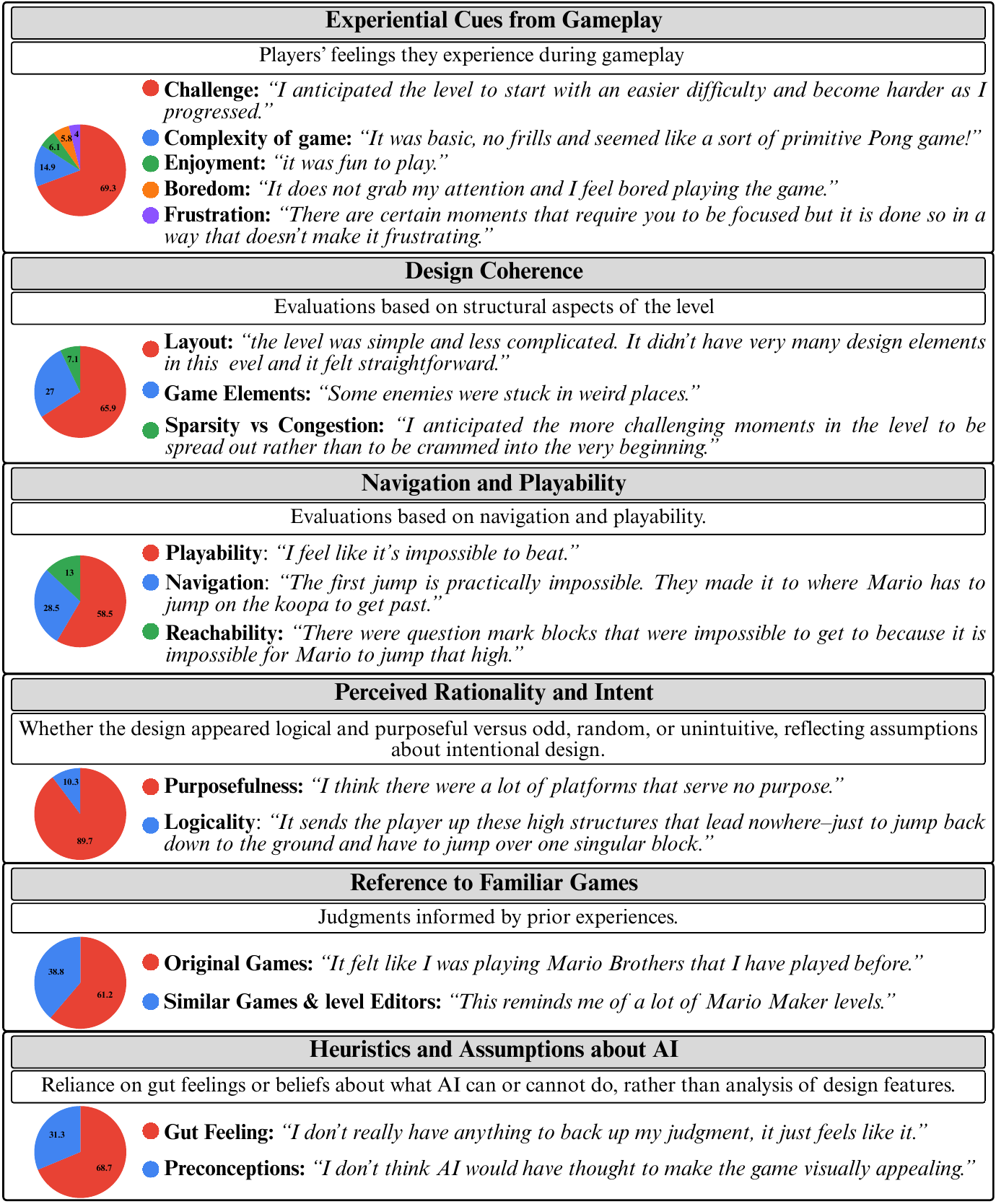}
\caption{Theme summary for strategies players used to distinguish AI-generated content, including codes associated with each theme and illustrative quotes. Pie plots show the distribution of code appearances within each theme.}
\Description{Themes used by players to identify AI-generated content: Experiential Cues from Gameplay (most repeated code: ``Challenge''), Design Coherence and Playability (most repeated code: ``Playability''), Perceived Rationality and Intent (most repeated code: ``Purposefulness''), Reference to Familiar Games (most repeated code: ``Original Game''), and Heuristics and Assumptions about AI (most repeated code: ``Gut Feeling''). Each codes come with a supporting illustrative participant quote.}
\label{fig:theme_summary_strategies}
\end{figure*}

We conducted a thematic analysis of 852 open-ended responses in answer to the question \textit{``What did you base your decision on? What signs or indicators guided your decision?''} Our initial coding produced 18 codes. Through iterative clustering and discussion, these were organized into 5 higher-level themes capturing how participants distinguish between AI- and human-generated game content (Figure \ref{fig:theme_summary_strategies}). We kept the codes and themes framed as strategies that players use to differentiate between AI-generated and human-created levels, without attaching them to a fixed outcome (AI-generated vs. human-created). This was an intentional choice, since interestingly, we observed that the very same justifications were often used to support opposite judgments about creator.

\subsubsection{\textbf{Themes}} Here we discuss the themes resulting from the analysis of strategies.

\noindent{\textbf{Experiential Cues from Gameplay}} were an important factor that players used to judge whether a level was made by AI or by a human. Players judged their overall experience with the level based on how interesting the level was and how much fun they had. In some cases, players also pointed to more complex aspects, like ``memorability'' when evaluating their experience: \customquote{The level feels ``just fine'' but lacks memorable moments.} Many players focused on the balance between challenge and frustration. For example, one participant described a Mario level as: \customquote{Very complicated jumps and odd layout that challenges you but doesn't offer entertainment as much as frustration.} Similarly, in Sokoban, some participants found the levels to be overwhelmingly challenging: \customquote{It's just a little too frustrating to play. It feels like it's impossible to win the game.} Boredom was another dominant feeling appearing more in Sokoban levels, with players mentioning: \customquote{It's very basic in feel or look to it, with no real creative design to it.  Just boring to look at.}

\noindent{\textbf{Design Coherence}} was another major way participants judged authorship. Players paid close attention to the overall layout of the level, including the sparsities and congestions, the symmetries and asymmetries. As one participant explained: \customquote{The beginning felt very cluttered and like there was too much going on.} In contrast, another participant noted how empty a section felt: \customquote{There are large areas of emptiness that drag down the immersion and quality of the level.}

Players also examined how elements such as enemies, coins, blocks, and pipes (in Mario) or boxes and targets (in Sokoban) were arranged. Sometimes placement felt deliberate: \customquote{You end up dying because of perfect enemy placement at the end of the jump. Seems like someone just wanted the level to be hard.} Other times it felt random: \customquote{There are enemy encounters that also seem random in that they don't pose a hazard because they resolve themselves by dying without interaction.}

\noindent{\textbf{Navigation and Playability}} Navigation and playability was another key concern. Players noticed when movement felt smooth versus when the level design made it awkward. They pointed out unreachable areas or spots where characters could get stuck. For example: \customquote{It felt very oddly designed, with the question mark blocks stacked so you couldn't actually get them.} or \customquote{Had blocks before the start for no reason and lots of wooden blocks high in air and impossible to get to}.

In some cases, levels were structured in a way that participants doubted whether they were even solvable. As one player put it in Mario: \customquote{This level was impossible. The gap was too big.} Similarly, in Sokoban: \customquote{I could not figure out a way to get the fourth block onto the target, and I'm uncertain that it is even possible.}

\noindent{\textbf{Perceived Rationality and Intent}} also shaped judgments. Players looked for whether design choices felt logical and purposeful, or instead odd and unintuitive: \customquote{The placement of coins and coin blocks seemed somewhat random and a little illogical in places to me.}

\noindent{\textbf{Reference to Familiar Games}} was another strategy. Some participants compared levels to the original game or similar games they remembered from playing. For example: \customquote{For one, I feel like I remember this level from playing as a kid.} Many also referenced their experiences with level editors like Mario Maker: \customquote{Have seen stuff like this in Super Mario Bros. Maker game.} or \customquote{The level has a lot of precision jumps. This reminds me of a lot of Mario Maker levels.}

\noindent{\textbf{Heuristics and Assumptions about AI}}  were used by players to make judgments about AI including what they thought AI could or could not create. For instance: \customquote{It feels very old in style and pixelated. That made me assume that it was made by a human.} or \customquote{It feels too complex for AI.}. In more extreme cases, some didn't have a clear idea of any automated level generation system and assumed older games must be human-made simply because of when they were released: \customquote{I think this is not AI because this game has been around for a long time, before AI.}. In other cases, participants mentioned they were simply guessing or relying on their gut feeling.

Looking across the themes that emerged in players' strategies for identifying the true origin of levels, it is clear that more strategies were readily applied to Mario levels than to Sokoban levels. For example, within Design Coherence and Playability, codes such as sparsity versus congestion, placement of game elements, and navigation or reachability were mentioned more often in Mario than in Sokoban. Similarly, judging purposefulness was more straightforward in Mario levels, while participants referred to this less often in Sokoban. This pattern aligns with Section \ref{sec:RQ1} results, where participants reported feeling generally less confident when judging Sokoban levels, since they had fewer cues available to guide their decisions.

\subsubsection{\textbf{Fallible Strategies}}
Even more interesting than identifying the strategies participants used to judge whether a level was AI-generated or human-created was seeing how often these strategies proved double-edged. The very same reasoning that led one person to mark a level as AI-generated was used by another to claim it must be human-made.

For example, when it came to \textbf{frustrating levels}, some participants felt that only humans would deliberately design super challenging levels to frustrate players: \customquote{human players are very capable of creating ``Hell'' type levels where the main goal is to frustrate players with impossible challenges, which actually is fun.} Others argued that frustration itself was a marker of poor AI design: \customquote{I feel like an AI doesn't really care if a human is frustrated trying to play a game since it has no concept of that.}

The same contradictions appeared with \textbf{easy levels}. Some saw simplicity as evidence of intentional human design, such as an introduction level: \customquote{It was a very simple game, but I do think that means it was created by a human for a starting level or as an introduction to the game.} Yet others felt simplicity was a sign of AI that is not capable of creating a more complex game: \customquote{It seemed like a very basic level with minimal logical puzzle solving involved. I believe AI probably made this.}

This ambiguity also applied to \textbf{challenging levels} that felt impossible to complete. Some thought impossibility was an AI flaw: \customquote{I find the level to be impossible to complete. I think this was made by an AI because it wouldn't understand the end goal of the game.}. Others believed unsolvability was more likely a novice human error: \customquote{This doesn't look like it can be solved. A human can make that mistake designing this, while AI is usually spot on.} In some cases, the same Mario level with a hard jump led to very different opinions. For some, the high level of challenge was an indicator of AI: \customquote{The first jump is practically impossible. They made it to where mario has to jump on the koopa to get past. That is something a human doesn't think of.} While the others saw the complicated mechanic as human intent: \customquote{It's hard but not impossible. I have an idea of what I'm supposed to do to make the jump (jump off the flying koopa) and it's the tricky kind of thing a human would put into the design of a level.}

Likewise, sometimes \textbf{odd design choices} were interpreted as mistakes or design choices. One participant explained the level with an enemy very close to where the player spawns as a human error: \customquote{It feels like there is human error where Mario dies each time before I can do anything.}. Others saw the exact level as an error only AI could make: \customquote{The level begins with an enemy immediately killing the player character. The level is set up in such a way that you have to be pressing the move key as soon as the level loads in order to escape immediate game over. No human would ever program a game in this manner.}. Interestingly, the same level was also perceived as an intentional joke (rather than a mistake) by some people: \customquote{It is impossible to complete the level as an enemy is placed directly beside the starting point. It feels like the level was created to troll others, which feels human to me.}. Similarly, another unusual feature of the level (an enemy falling off a cliff) was interpreted in completely opposite ways: for some it signaled pointless design that AI creates: \customquote{So many pointless elements like Koopas that just fall into the void.}, while again for others it was read as an intentional joke made by a human: \customquote{The goombas falling off the cliff at the beginning felt like an intentional joke by a person.}


All in all, these contradictions show how weak the strategies were. The same design feature could just as easily be used to argue for AI authorship as for human authorship, suggesting that judgments often rested on subjective interpretations rather than reliable cues.
\begin{table*}[h!]
\centering
\begin{tabular}{lcccc}
\toprule
 & \multicolumn{2}{c}{\textbf{Perceived as AI (n=428)}} & \multicolumn{2}{c}{\textbf{Perceived as Human (n=424)}} \\
\textbf{Metric} & Mean (SD) & Median [IQR] & Mean (SD) & Median [IQR] \\
\midrule
Enjoyment/Fun & 2.92 (1.33) & 3 [2-4] & 3.72 (1.20) & 4 [3-5] \\
Difficulty/Challenge & 3.88 (1.27) & 4 [3.8-5] & 3.65 (1.32) & 4 [3-5] \\
Frustration/Negative Affect & 3.60 (1.37) & 4 [2-5] & 2.84 (1.43) & 3 [2-4] \\
Novelty/Surprise & 2.65 (1.35) & 2 [1-4] & 2.62 (1.26) & 3 [1-4] \\
Aesthetics/Design Quality & 2.70 (1.23) & 3 [2-4] & 3.57 (1.17) & 4 [3-4] \\
\bottomrule
\end{tabular}
\caption{Descriptive statistics for five experience metrics by perceived source of creation. All metrics were measured on a 5-point Likert scale (1 = Strongly disagree, 5 = Strongly agree). IQR = Interquartile range.}
\label{tab:descriptive_stats}
\end{table*}

\subsection{RQ4: Is there a link between players' beliefs of the origin of the level and how they rate their experience with the game levels?}
\begin{figure*}[h]
    \centering
    \begin{minipage}{0.9\linewidth}
        \centering
        \includegraphics[width=\linewidth]{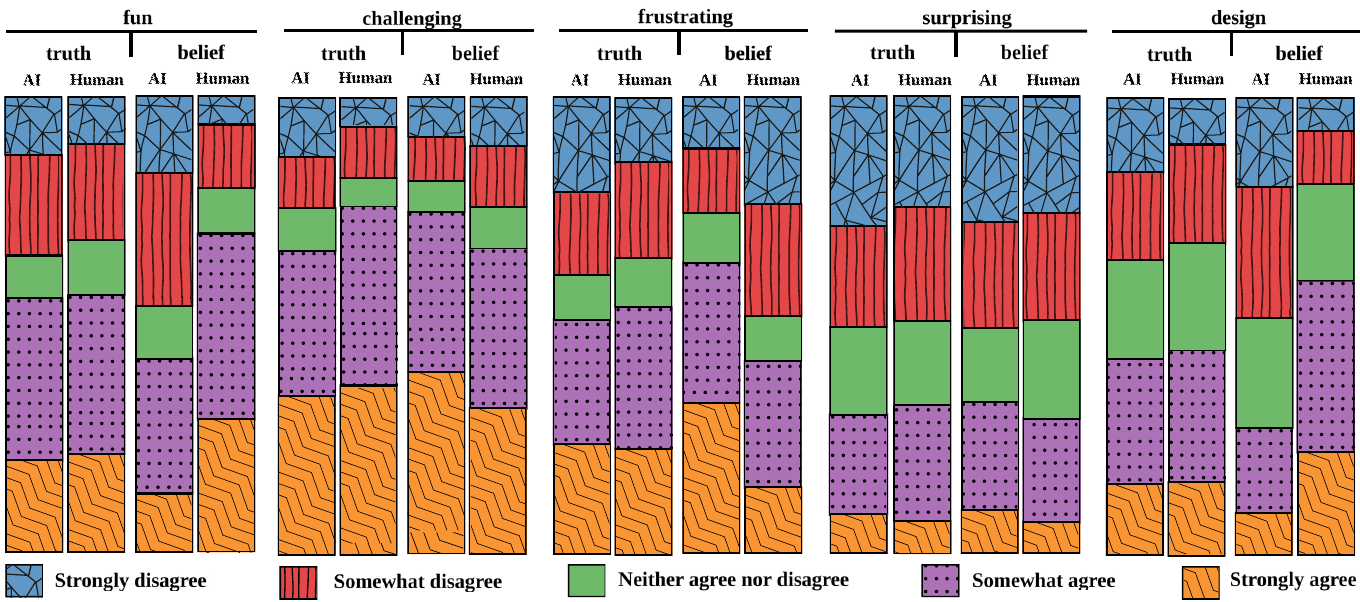}
        \caption{Stacked bar charts showing player ratings of five experience metrics (\metric{fun}, \metric{challenging}, \metric{frustrating}, \metric{surprising}, \metric{design}). For each metric, separate bars show ratings for truly AI-generated or human-made levels~(marked as ``truth'') and for levels believed to be AI- or human-made~(marked as ``belief''). Bar segments represent proportions of Likert responses from ``strongly disagree'' to ``strongly agree.''}
        \Description{Stacked bar charts display player ratings across five experience metrics:\metric{fun}, \metric{challenging}, \metric{frustrating}, \metric{surprising}, \metric{design}. Each metric includes two bars—one for the actual origin (marked as ``truth'') and one for the perceived origin (marked as ``belief'') of the game level (either could be AI or human-made). Bar segments show Likert-scale proportions from ``strongly disagree'' to ``strongly agree.'' When grouped by true origin, rating differences are modest, but when grouped by belief, differences are sharper.}
        \label{fig:full_experience}
    \vspace{0.5cm}
    \end{minipage}\hfill
    \begin{minipage}{0.9\linewidth}
        \centering
        \includegraphics[width=\linewidth]{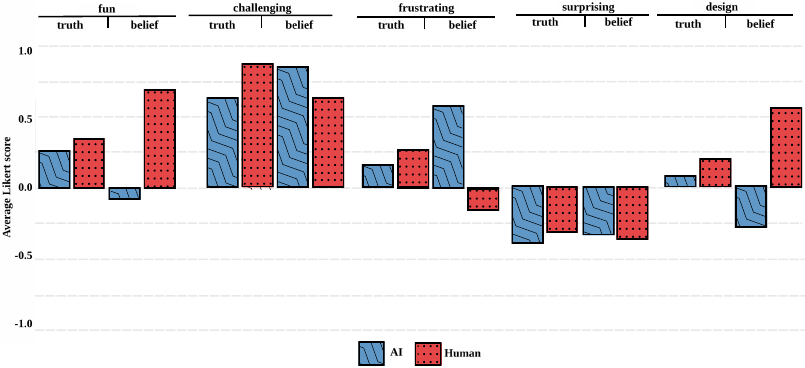}
        \caption{Mean player ratings across five experience metrics (\metric{fun}, \metric{challenging}, \metric{frustrating}, \metric{surprising}, \metric{design}), grouped by the true source of creation of the level and by participants' belief about the origin.}
        \Description{Grouped bar chart shows mean player ratings for five experience metrics(\metric{fun}, \metric{challenging}, \metric{frustrating}, \metric{surprising}, \metric{design}) split by true origin and belief about level origin. Blue waved bars represent ``AI'' and red dotted bars represent ``human''. Ratings grouped by actual origin (truth) show modest differences, while belief-based groupings show sharper contrasts.}
        \label{fig:metrics_average}
    \end{minipage}
\end{figure*}
We computed the mean Likert score for each metric by mapping responses onto a numeric scale from –2 (strongly disagree) to +2 (strongly agree) and then averaging across responses. The table \ref{tab:descriptive_stats} shows the descriptive statistics for all metrics by perceived source of creation. Also, Figure \ref{fig:metrics_average} visualizes the mean Likert scores and  Figure \ref{fig:full_experience} visualizes the original distribution of Likert responses without this transformation.

When grouped by the true creator (AI vs. Human), differences in mean ratings were relatively modest. However, when grouped by belief, the differences became much sharper, hinting there may be a relationship between players' experiences and their assumptions about the creator, rather than by the objective qualities of the levels themselves. Our statistical analysis confirms this observation. Across \metric{Fun}, \metric{Challenge}, \metric{Frustration}, and \metric{Design}, participants' beliefs about whether a level was AI- or human-made consistently outweighed the true creator as predictors of experience ratings.

For \metric{Fun}, players rated levels significantly more enjoyable when they believed the level was human-made (\stat{$\beta$ = 1.54, SE = 0.16, z = 9.52, p < .001}). The true source of creation was not significant (\stat{$\beta$ = 0.27, SE = 0.20, z = 1.35, p = .18}). Generally, Sokoban levels were rated significantly less fun than Mario levels (\stat{$\beta$ = -0.71, SE = 0.20, z = -3.53, p $<$ .001}).

\metric{Challenge} ratings were also linked to belief. Levels that were believed to be AI-generated were rated more challenging (\stat{z = -2.41, p < 0.015}). The true creator (\stat{p = 0.28}) and game type (\stat{p = 0.19}) were not significant predictors.  

\metric{Frustration} ratings depended strongly on belief. When participants thought a level was AI-generated, they rated it as more frustrating (\stat{$\beta$=-1.17, SE = 0.16, z = -7.445, p < .001}). Neither the true creator ($p = 0.38$) nor the game type (\stat{p = 0.37}) was significant.  

Perceptions of \metric{Design} and aesthetics were also more associated with belief than with reality. Levels believed to be human-made were rated significantly more aesthetically pleasing (\stat{z= 10.480, p < .001}). The true creator did not have a significant effect (\stat{p = 0.087}), but the game had, with Sokoban levels rated lower in design than Mario levels (\stat{z = -3.373, p < 0.001}).

Participants' ratings for \metric{Surprise} were not related to their belief about the creator (\stat{$\beta$ = 0.01, SE = 0.15, z = 0.05, p = .96}) or the true creator itself ($\beta = 0.17$, $SE = 0.23$, $z = 0.74$, $p = .46$). Instead, the only significant predictor was the type of game as Sokoban levels were rated significantly less surprising compared to Mario levels (\stat{$\beta$ = -1.14, SE = 0.23, z = -4.88, p < .001}).

Overall, across all experience metrics, \textbf{participants' beliefs} about the origin of the level were more strongly associated with their ratings than the \textbf{true source} of creation. Also, the generally lower scores of Sokoban levels on \metric{Fun}, \metric{Design}, and \metric{Surprise} are consistent with the comments in RQ2 (Section \ref{sec:RQ2}), where participants described these levels more as boring and less as fun.
\subsection{RQ5: Do players have different attitudes toward PCG and generative AI? How does this prior relate to their reported experiences with the game levels?}
\begin{figure*}[!h]
\centering
\includegraphics[width=0.98\textwidth]{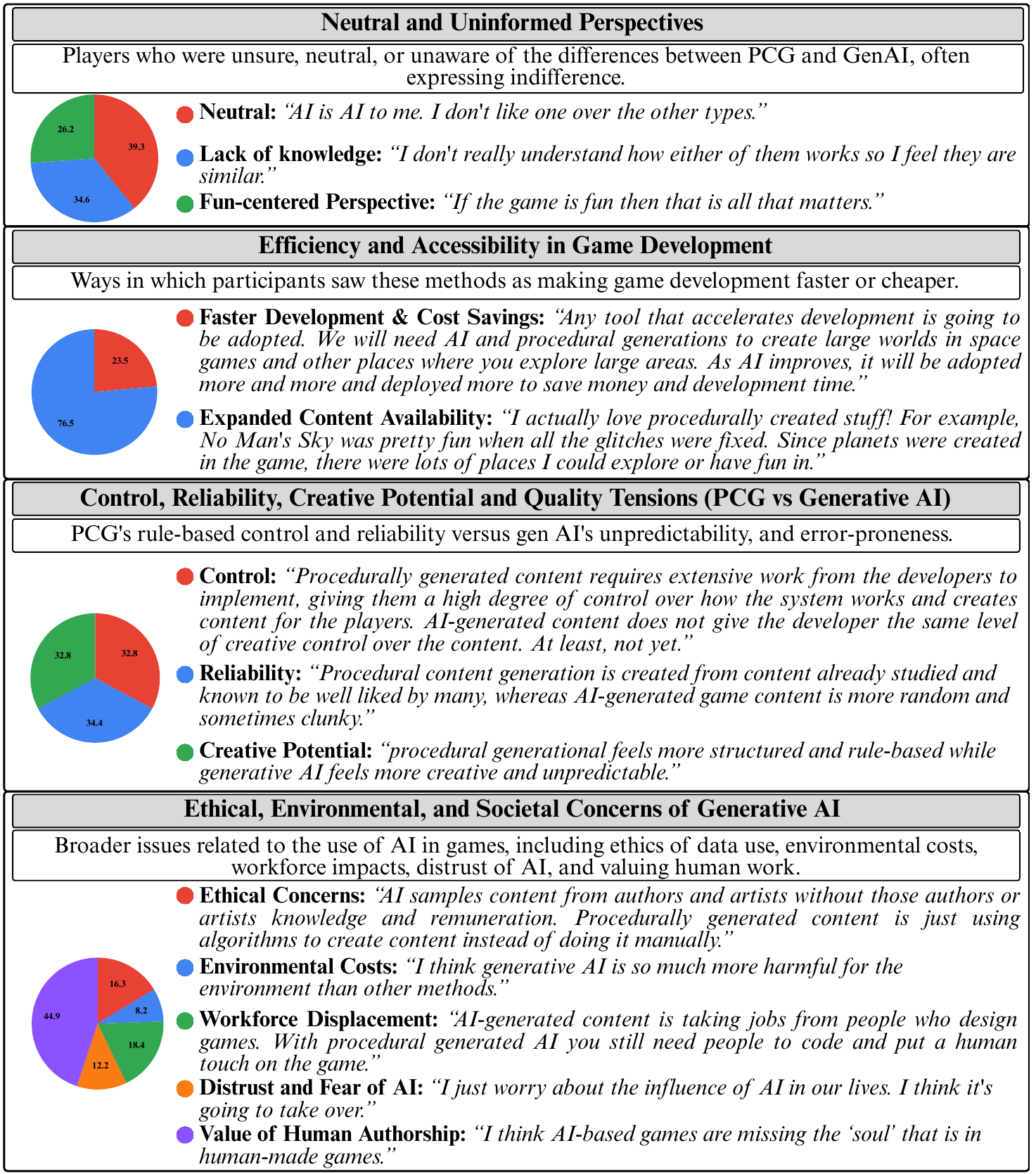}
\caption{Theme summary for views on generative AI and PCG in games, including codes associated with each theme and illustrative quotes. Pie plots show the distribution of code appearances within each theme.}
\Description{Themes present in players' views on generative AI and PCG in games. Themes include Neutral and Uninformed Perspectives (most repeated code: ``Neutral''), Efficiency and Accessibility in Game Development (most repeated code: ``Expanded Content Availability''), Control, Reliability, Creative Potential and Quality Tensions (most repeated code: ``Reliability''), and Ethical, Environmental, and Societal  (most repeated code: ``Value of Human Authorship''). Each codes come with a supporting illustrative participant quote.}
\label{fig:theme_summary_bias}
\end{figure*}
We compared participants' attitudes toward procedural content generation (PCG) and generative AI, both measured on a 5-point Likert scale ranging from Very negative to Very positive. A Pearson's chi-square test indicated that the distributions of responses differed significantly between PCG and generative AI (\stat{$\chi^2(16)$ = 473.71, p < .001}). The result was also consistent with a non-parametric Wilcoxon signed-rank test on numeric-coded scales, which also showed a highly significant difference (\stat{V = 47919, p < .001}). Together, these results indicate that participants' reported attitudes toward PCG and generative AI were systematically different, with more negative views expressed toward generative AI (Figure \ref{fig:bias}).

Participant' attitudes toward generative AI played an important role in shaping their experience. \metric{Fun} ratings were significantly higher among participants with more positive views of generative AI (\stat{z= 3.247, p = 0.0011}). In contrast, attitudes toward PCG did not significantly predict fun, though there was a weak trend (\stat{p = 0.060}). \metric{Design} ratings were also positively related to participants' views of generative AI, with more favorable attitudes predicting significantly higher aesthetic scores (\stat{z = 2.391, p = 0.0168}). Neither PCG nor generative AI views had significant effects on the rest of the perceived factors.

To investigate this further, we also conducted a thematic analysis of 142 open-ended responses in answer to the question \textit{``Do you feel differently about AI-generated game content and procedurally generated content? Why?''} Our initial coding produced 14 codes. Through iterative clustering and discussion, these were organized into 4 higher-level themes capturing the origin of participants' views on generative AI and PCG in games~(Figure \ref{fig:theme_summary_bias}).

\subsubsection{\textbf{Themes}} Here we discuss the themes resulting from the analysis of AI and procedurally generated content.

\noindent\textbf{Neutral or indifferent views},
expressed by some participants, seeing little distinction between PCG and generative AI, or admitting they did not understand how the methods worked. For them, the central point was whether the game itself was enjoyable: \begin{quote}
    \customquote{I regularly make ChatGPT and Copilot act as a game master and I give it a background story to build the world upon and I give it certain rules to follow, like game mechanics, random battle encounters, loots, things like that. It's very fun because I can just get in the game that never ends (until I tell it to end).  I love games, whether humans or AI made them.}
\end{quote}

Others described both methods as similar or said that ``AI is AI'' without differentiating: \customquote{For someone who isn't an expert on AI, when you tell me both of these things are computer-generated or AI, it all seems the same to me.} This theme highlights that not all players approach content generation with strong priors and for some the technical origin of the content was irrelevant.

\noindent\textbf{Making game development faster, cheaper, and more accessible} was another common view about both PCG and Generative AI. Participants often framed these methods as tools for scaling up large game worlds or expanding the amount of content: \customquote{They are both types of games that if done correctly can provide an immersive, fun experience.  I like variety in my games.}

\noindent\textbf{Comparing PCG with generative AI} was done in the most interesting and well-justified responses from participants who had enough knowledge of the details and differences between these methods to make informed judgments. PCG was often described as reliable and rule-based, that gives developers more creative control. In contrast, Generative AI was described as unpredictable, error-prone, and less structured: 
\begin{quote}
    \customquote{Procedural content generation is rule-based, so the results are predictable within certain limits and feel consistent with the game's design. Generative AI, on the other hand, is more open-ended and can create unexpected, highly varied results based on prompts, which can be exciting but also less controlled. The difference is in predictability versus creativity.}
\end{quote}

Some participants had a positive take on generative AI for its creativity and unpredictability, suggesting it might bring more novelty to games: \customquote{I believe that procedurally generated game content leads to a barrage of the same types of experiences repeating themselves with little variation, whereas AI-generated game content has the potential to be more of a unique experience.} While others believed generative AI does not produce original content and felt negative about it: \customquote{Generative AI is only as strong as the datasets used to create it, which means you're automatically getting derivative and unoriginal content.}

\noindent\textbf{Broader issues about Generative AI} were frequently raised  by participants. Ethical worries included the use of unlicensed data and the lack of credit to artists: \customquote{Procedurally generated game content still had to be programmed by someone and took actual work by actual people. Generative AI content is regurgitated slop stolen from real artists. They are not the same.}

Others expressed concerns about the environmental costs of training AI models, workforce displacement in creative industries, and a more general distrust of AI. Participants also frequently emphasized their appreciation of game design as a human creation, noting that they value the unique ``human touch'' in games. As one participant explained: \customquote{It is fine to use AI, but there is a different aspect when it is human-made that almost shows more effort and I get more enjoyment out of it, even if it isn't the highest quality.}

These perspectives show that judgments about Generative AI were not only about gameplay but were tied to wider societal debates. Of course, these are not the only ethical and sociological issues raised by generative AI (those also include other debates around misinformation, bias, cultural appropriation, etc). The mentioned items were the concerns that appeared most clearly in our thematic analysis.
\section{Discussion}

\subsection{Players use double-edged heuristics of ``Human-likeness''}
Players draw on a wide range of strategies to decide whether a level is AI- or human-made and interestingly, they align with patterns reported in other domains of AI-content detection. For example, just as readers look for subtle cues of errors or odd phrasing in AI text~\citep{nabata_evaluating_2025}, our players inspected playability and odd design choices. Similarly, prior studies on image-based Turing tests show that people look for logical consistency~\citep{lu_seeing_2023} and anomalies in details~\citep{bray_testing_2023}, which resonate with our players' sensitivity to unreachable areas, wasted space, or off-balance layouts.

More importantly, we observed that the very same cues could be double-edged. Odd design choices were read as either deliberate human trolling or as careless AI design; simplicity in game design was taken as evidence of either human intention or machine limitation. This also mirrors previous work findings in other domains like text (grammatical errors in text can be read as either human error or ``bad'' AI~\citep{jakesch_human_2023}), or visual arts (irregularities in drawings can be seen differently depending on whether they are attributed to humans or AI~\citep{daniele_what_2021}).

Taken together, these findings suggest that ``human-likeness'' is not an objective threshold that can be engineered into content, but a shifting construct defined by players' own expectations, biases, and interpretive strategies. In games, this means that players may disagree not only on whether a level feels human- or AI-made, but also on what qualities even count as evidence of humanness. This helps to explain why more experienced players do not necessarily improve at distinguishing AI-generated content from human-created content. They become more confident in their judgments as their experience increases, but their accuracy remains close to chance.
\subsection{Perceived creator, not truth, shapes players' experience}
Players cannot reliably tell AI-generated levels from human-created levels. This result aligns with earlier Turing-style studies from a decade ago on Super Mario Bros.~\cite{reis_human_2015} and Sokoban~\citep{taylor_comparing_2015}, where players also failed to detect the creator even when classical search-based methods were used. Yet the levels believed to be AI-generated were rated lower on experiential dimensions such as \metric{Fun} and \metric{Challenge}, and more \metric{Frustration} and \metric{Design}, showing that AI carries a negative bias. Similar dynamics have been observed in cooperative play, where perceived partner identity matters more than actual performance. For example, \citet{merritt_choosing_2011} found that players enjoyed and preferred ``presumed human'' teammates over identical AI-controlled teammates. Likewise, \citet{ragot_ai_2020} found that artworks believed to be AI-made were rated lower in beauty, meaning, and liking compared to the same works attributed to humans.

But more importantly, this negative bias appeared \textbf{without any priming or explicit disclosure} in our study, suggesting that spontaneous judgments about creator can be just as powerful as judgments shaped by labels. Players' experiences were more closely associated with what they believed about a level's creator than with its actual creator. 
\subsection{Ambiguous signals create a lemons dynamic}
\citet{de_when_2023} used the ``lemons problem'' in economics~\citep{akerlof_market_1978} to show that when people cannot tell which content is real~(human-authored) and which is not real~(AI-generated), they may start distrusting everything. In games, players also cannot reliably tell whether content is AI- or human-made. Current labels (``AI content,'' ``procedural content'') are overly broad, grouping together very different practices such as brainstorming ideas, code and game logic, marketing materials, narrative generation, and visual asset creation. These vague labels risk priming negative reactions before play even begins~\citep{ragot_ai_2020, samo_artificial_2023, zhu_human_2024, grassini_understanding_2024}, leading some developers to avoid disclosure altogether. Yet our findings show that even without disclosure, when players speculate about AI involvement the same bias emerges. This creates a ``lemons'' dynamic, where ambiguous signals and perception bias lower perceived quality regardless of the actual content.
\subsection{Generative AI carries stigma beyond play}
Our study shows that players reported systematically different attitudes toward PCG and generative AI. While some participants expressed indifference and emphasized enjoyment as the ultimate goal, overall responses were more negative toward generative AI. This aligns with prior research findings that generative AI reshapes creative practices in both promising and worrying ways~\citep{vimpari_adapt_2023}, creating tensions between technical affordances and ethical commitments~\citep{boucher_resistance_2024, panchanadikar_golden_2024}. Our study extends these insights by showing that such uncertainty is also present among players.

Importantly, participants' concerns also extended beyond gameplay to debates about authorship, environmental costs, and labor conditions. These concerns resonate with wider discussions in the games industry. Like the interns in \citet{boucher_resistance_2024} study, they questioned the legitimacy of AI-generated art, pointing to unresolved ethical issues around training data that fails to acknowledge or human creators. Similar to indie developers in the \citet{panchanadikar_solo_2024} and \citet{vimpari_adapt_2023} studies, they worried that generative AI could erode creativity and fair labor. These perspectives suggest that players are not passive consumers and that they play an active role in the conversations around AI in games.

Aligned with our observations on players' views on PCG, \citet{panchanadikar_solo_2024} observed that generative AI unsettles established practices in creative industries, whereas PCG is now more normalized as a design material with predictable boundaries. Yet we should remember this normalization did not occur automatically. Early PCG was often criticized as repetitive or soulless, from the dungeons of roguelikes to the sparse planets of No Man's Sky's initial release~\citep{ruffino_19_2024}. As PCG got better integrated into design practices, people's perceptions started to change, and they came to value it for the replayability and scale it provides. Similarly, the stigma around generative AI may lessen if the broader ethical issues are addressed. Yet such a change will not occur automatically and it requires confronting existing issues. One viable path is greater transparency achieved not through broad labels but through nuanced disclosures.
\subsection{Nuanced disclosure as a path to trust}
Nuanced disclosures that explain \emph{how} and \emph{why} AI was used offer a promising way forward. Prior work shows that developers welcome AI for tasks like ideation and prototyping but grow uneasy when it extends into authorship-intensive areas such as asset creation~\citep{alharthi_generative_2025}. Players also draw similar boundaries. For some people, using AI in narrative design may be fine, while its use in asset generation is more problematic. Providing contextual details about where and why AI is used both helps players to make informed judgments and protects developers from being mischaracterized by broad and vague labels. For indie developers especially, such disclosures are more necessary to support them in experimenting with new tools.
\subsection{Limitations and future directions}
Our study has several limitations. It was restricted to two short, 2D tile-based games (Super Mario Bros. and Sokoban). While common benchmarks in PCG research, it remains unclear how results generalize to other genres. Participants also only played short levels in our study. Although this design allowed us to collect data across many conditions, it does not capture the richer experiences of longer gameplay that include extended immersion, storylines, and evolving mechanics.

Our survey design relied on single-item Likert ratings of experiential constructs. Although efficient for online data collection, single-item measures may not capture the full complexity of these constructs. Future research could adopt established multi-item instruments like GEQ and GUESS to provide more robust and reliable measures. Our thematic analysis was conducted on relatively brief open-ended answers. While these responses offered valuable insights into players' strategies and views on PCG and generative AI, they lacked the depth of longer interviews or think-aloud protocols. Additionally, our sample was drawn from online crowd participants, whose gameplay habits and motivations may differ from dedicated player communities. Also, although this population reflects the actual distribution in the 
general population (where most gamers don't have game design experience), the under-representation of experienced game designers limits our statistical power to detect experience-related effects. Future work with targeted recruitment of level designers could provide stronger tests of expertise effects. 

Our study focused on players' subjective perceptions. This means we ensured that AI-generated levels were playable and aesthetically acceptable to avoid trivial distinguishability, but did not include objective structural or design-level comparisons between AI-generated and human-generated levels. Future work could incorporate systematic structural analysis (e.g., metrics for spatial complexity, challenge curves, or aesthetic patterns) to disentangle the effects of objective level characteristics from subjective perceptual biases. 
Our design was observational: we observed associations between perception about the creator and experience, but we cannot establish causality. This means that the causal direction may run in two ways. It could be the expectation shaping experience, meaning that when players suspect ``AI involvement'' they may focus more on flaws. Alternatively, it could be that players infer the creator from their experience (e.g., ``this was frustrating, so it must be AI''). Clarifying this causal pathway (through controlled disclosure manipulations based on player perceptions) would be a particularly interesting and valuable direction for future research.

Despite the limitations, our design provided interesting empirical results. Future work can extend this work by exploring additional genres, incorporating longer play sessions, and recruiting participants with higher experience with game design.

\section{Conclusion}
This paper examined how players' perception of the creator of game levels is associated with their experience of the gameplay. We conducted a mixed-method study that included a Turing-style classification task, quantitative player experience measures, and qualitative thematic analysis of open-ended responses. We found that players' accuracy in distinguishing AI-generated from human-designed levels was close to random guessing. We analyzed the strategies players use for judging ``human-likeness'' and we observed how these strategies are subjective and fallible. Our findings also show that players' perceptions about the authorship of levels were strongly associated with how they evaluated them. Levels that players thought were made by humans were usually seen as more fun and visually appealing. In contrast, levels they believed came from AI were more often described as frustrating or overly difficult. When players were asked about their views on AI-content in games, their opinions were more negative on generative AI than with PCG, with some comments addressing the reliability and ethical risks that come with generative AI. Finally, we discussed the need for more transparency on how AI is used in games, and we raised new questions about ownership and trust of AI-generated content in games for future work.
\begin{acks}
We thank Aria Masoomi for helping with thematic analysis coding. We also thank CHI Crunch group at Ghost Lab, Northeastern University, and Casper Harteveld for providing feedback and encouragement during the development of this paper.
\end{acks}
\bibliographystyle{ACM-Reference-Format}
\bibliography{refs}

\appendix
\section{Appendix}
\begin{figure*}[t!]
    \centering
    \includegraphics[width=1\linewidth]{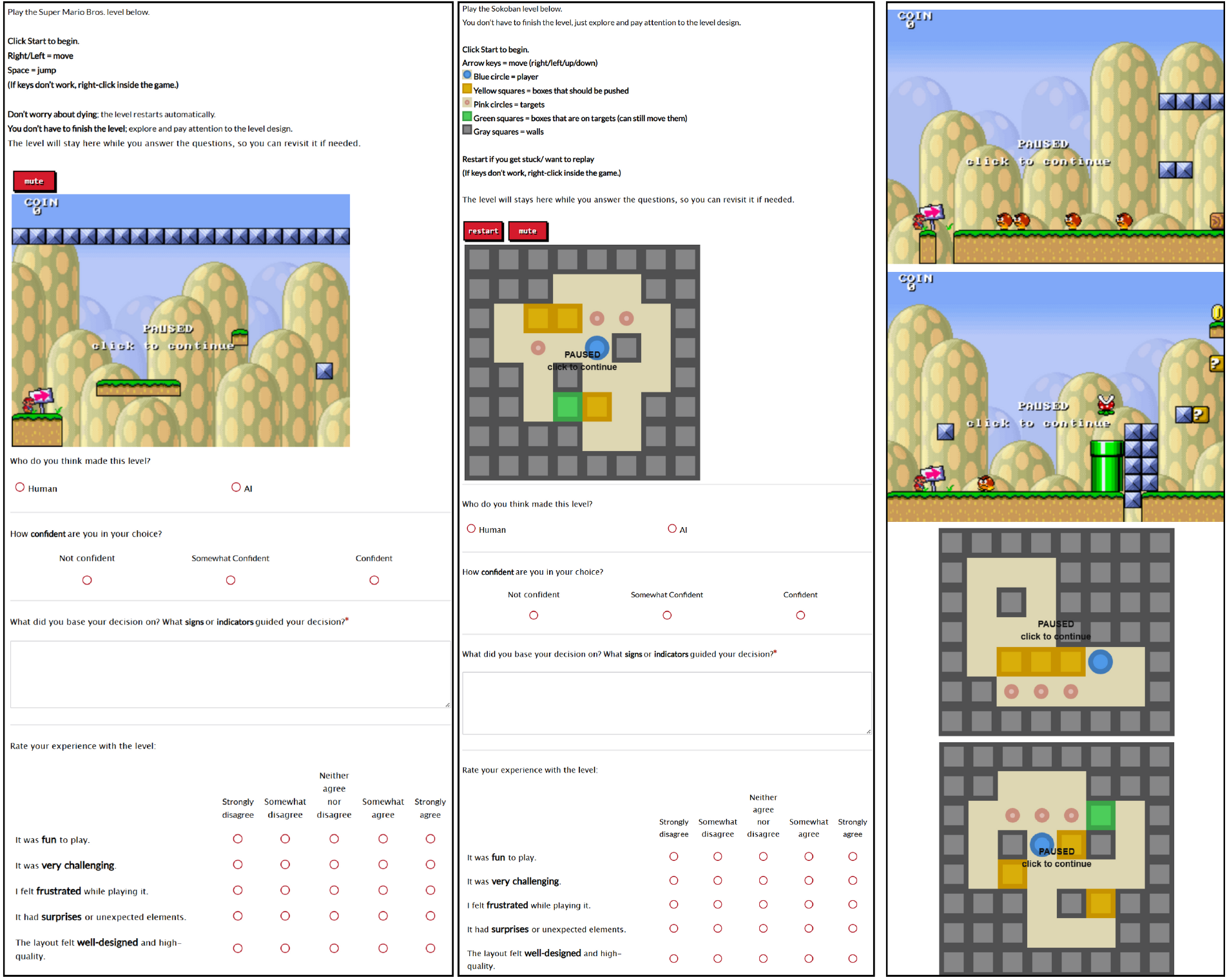}
    \caption{Screenshots of the survey interface as seen by one participant. The complete screenshot with all instructions and questions is shown for the first Super Mario Bros. and Sokoban levels. For subsequent trials to conserve space, only the game level screenshots are added. Note that unlike Sokoban, Super Mario Bros. levels are not completely visible in the player's viewport, and horizontal movement is needed to see the rest of the level.}
    \Description{Screenshots of an actual survey presented to a participant. Each trial includes a game and the follow-up questions.}
    \label{fig:survey}
\end{figure*}
\subsection{AI generation Method Selection}\label{app:lit_rev}
In order to select level generation approaches we conducted a search of relevant literature. We were trying to find papers in the domain of PCG that exactly included the generation of Super Mario Bros. and Sokoban levels. We decided to find papers that included a codebase of their implementation so that we could exactly use the model provided by the authors to generate random models. To do so we conducted a literature search with the keyword string \texttt{“(mario OR sokoban) AND 
(pcg OR procedural OR generative OR PCGML OR PCGRL
OR constraint OR Constraint-Based OR WFC OR ‘Wave Function Collapse’ OR WaveFunctionCollapse
OR Markov OR Bayesian Networks OR latent
OR Supervised OR
OR Neural OR Autoencoder OR recurrent OR RNN OR LSTM OR "Long Short-term Memory"
OR transformer OR LLM "Large Language" OR GPT
OR convolutions OR Convolutional OR CNN OR Adversarial OR GAN OR Variational OR VAE
OR Reinforcement OR RL OR MDP OR Q-learning
OR Diffusion)”}.
This keyword string was selected based on all approaches mentioned in different sections of~\citep{yannakakis_procedural_2025}. Searching in IEEE Xplore, ACM DL, and AAAI databases resulted in 446 records (227 IEEE Xplore, 168 ACM DL and 51 AAAI). Then we removed 73 duplicate records and 10 ineligible records (marked by automation tools), resulting in 361 unique papers for screening. Twenty papers were excluded at the title/abstract screening stage as they clearly did not focus on level generation for our target games despite containing our search keywords.  After applying our inclusion criteria (full published papers focusing on Super Mario Bros. or Sokoban level generation) and exclusion criteria (removing short papers/abstracts (44), review papers (11), papers not covering our target games (167), mixed-initiative/repair/blending methods (6), and crucially, papers without available code/artifacts (93)) we identified 20 papers with reproducible implementations.

From the 20 papers meeting our inclusion criteria, we selected 6 representative models (3 for SMB, 3 for Sokoban) following a paired-method approach to keep methodological consistency across both games. We first identified method categories that had implementations available for both Super Mario and Sokoban, which excluded GAN-based, genetic, and grammar-based approaches that were only available for SMB generation. Paper~\citep{cooper_literally_2024} was excluded as it specifically generates unplayable levels.
This filtering process resulted in four method categories with representation in both games: constraint-based (3 papers:~\cite{cooper_sturgeon_2022},~\citep{cooper_sturgeon_2023},~\citep{cooper_sturgeon_2024}), reinforcement learning (5 papers:~\citep{wang_fun_2022},~\citep{shu_experience_2021},~\citep{wang_online_2022},~\citep{khalifa_pcgrl_2020},~\citep{racaniere_imagination_2017}), large language models (2 papers:~\citep{sudhakaran_mariogpt_2023},~\citep{todd_level_2023}). 
, and evolutionary methods (2 papers:~\citep{beukman_procedural_2022},~\citep{beukman_procedural_2022}).

For LLMs, only one implementation per game was available, so no choice needed to be made. For constraint-based and RL methods where multiple options existed, we selected based on citation count as a proxy for impact and reliability. We chose \citep{cooper_sturgeon_2022}~(34 citations) for Super Mario Bros. and \citep{cooper_sturgeon_2023}~(14 citations) for Sokoban from constraint-based methods. We also chose \citep{wang_fun_2022}~(16 citations) for Super Mario Bros. and \citep{racaniere_imagination_2017}~(795 citations) for Sokoban from  RL approaches. While \citep{khalifa_pcgrl_2020}~(243 citations) and \citep{shu_experience_2021}~(71 citations) had higher citation counts than \citep{wang_fun_2022}, they were excluded due to deprecated dependencies in their codebases that prevented reliable reproduction. Also, we excluded evolutionary approaches because both available papers used reinforcement learning as their core generation mechanism, which would have been redundant with our RL category.

\subsection{AI generation Process}\label{app:generation_resources}
To ensure reproducibility and transparency, we provide the exact resources used for each generation approach in our study. For approaches where trained models were available, we used these models to generate levels. For approaches where complete artifact datasets from the original work were available, we used these pre-generated levels.

\textbf{Trained Models:}

\begin{itemize}
    \item \textbf{Cooper et al. (2022)}~\citep{cooper_sturgeon_2022}: We used the model exactly as instructed in their GitHub repository\footnote{\url{https://github.com/crowdgames/sturgeon/blob/main/examples_basic.sh}} lines 52-70.
    
    \item \textbf{Cooper et al. (2023)}~\citep{cooper_sturgeon_2023}: We used the model exactly as instructed in their GitHub repository\footnote{\url{https://github.com/crowdgames/sturgeon/blob/main/examples_mkiii.sh}} lines 28-30.
    
    \item \textbf{Sudhakaran et al. (2023)}~\citep{sudhakaran_mariogpt_2023}: We used the trained MarioGPT model through the provided Colab notebook\footnote{\url{https://colab.research.google.com/drive/16KR9idJUim6RAiyPASoQAaC768AvOGxP}}. For this model, we used the prompts as specified in their Figure 1: 
    \begin{enumerate}
        \item ``many pipes, many enemies, little blocks, low elevation''
        \item ``no pipes, some enemies, many blocks, high elevation''
        \item ``many pipes, many enemies''
        \item ``no pipes, no enemies, many blocks''
        \item ``many pipes, no enemies, many blocks''
        \item ``Many pipes, no enemies, some blocks'' 
    \end{enumerate}
\end{itemize}

\textbf{Artifact Datasets:}

\begin{itemize}
    \item \textbf{Wang et al. (2022)}~\citep{wang_fun_2022}: We used the pre-generated level dataset from their repository\footnote{\url{https://github.com/SUSTechGameAI/MFEDRL/tree/master/exp_data/rand_playable_lvls}}
    
    \item \textbf{Todd et al. (2023)}~\citep{todd_level_2023}: We used their artifact dataset available online\footnote{\url{https://tinyurl.com/sokoban-todd}}
    
    \item \textbf{Racanière et al. (2017)}~\citep{racaniere_imagination_2017}: We used the extensive dataset of generated levels from the original work's repository\footnote{\url{https://github.com/google-deepmind/boxoban-levels}}
\end{itemize}

\subsection{Qualtrics Survey}
Figure \ref{fig:survey} shows screenshots of the survey interface as seen by one participant.

\end{document}